\newcommand{\be}{\begin{equation}}
\newcommand{\bea}{\begin{eqnarray}}
\newcommand{\ee}{\end{equation}}
\newcommand{\eea}{\end{eqnarray}}
\newcommand{\nn}{\nonumber}

\newcommand{\qa}{\alpha}
\newcommand{\qb}{\beta}
\newcommand{\qg}{\gamma}

\newcommand{\qd}{\delta}

\newcommand{\qe}{\varepsilon}

\newcommand{\qh}{\eta}

\newcommand{\qk}{\kappa}
\newcommand{\ql}{\lambda}
\newcommand{\qL}{\Lambda}

\newcommand{\qr}{\rho}
\newcommand{\qs}{\sigma}
\newcommand{\qS}{\Sigma}

\newcommand{\qF}{\Phi}

\newcommand{\qj}{\psi}
\newcommand{\qJ}{\Psi}
\newcommand{\qo}{\omega}
\newcommand{\qO}{\Omega}

\newcommand{\tr}{{\rm tr}\,}

\newcommand{\tri}{\triangle}

\newcommand{\dagg}{^{\dag}}

\newcommand{\fr}[2]{{\textstyle \frac{#1}{#2}}}

\newcommand{\EE}{\mathop{{\mathbb{E}}}}

\newcommand{\one}{{\mathbb 1}}
\newcommand{\sH}{{\sf H}}
\newcommand{\bits}{ \{0,1\} }

\newcommand{\cA}{{\mathcal A}}
\newcommand{\cB}{{\mathcal B}}

\newcommand{\cE}{{\mathcal E}}
\newcommand{\cF}{{\mathcal F}}

\newcommand{\cH}{{\mathcal H}}
\newcommand{\cI}{{\mathcal I}}

\newcommand{\cK}{{\mathcal K}}
\newcommand{\cL}{{\mathcal L}}
\newcommand{\cM}{{\mathcal M}}

\newcommand{\cO}{{\mathcal O}}

\newcommand{\cQ}{{\mathcal Q}}

\newcommand{\cS}{{\mathcal S}}
\newcommand{\cT}{{\mathcal T}}

\newcommand{\cX}{{\mathcal X}}

\newcommand{\pr}{{\rm Pr}}

\newcommand{\isdef}{\stackrel{\rm def}{=}}

\newcommand{\ket}[1]{| #1 \rangle}
\newcommand{\bra}[1]{\langle #1 |}
\newcommand{\inprod}[2]{\langle #1 | #2 \rangle}

\documentclass{article}
\usepackage{graphics}
\usepackage{graphicx}
\usepackage{url}
\usepackage{amsmath}
\usepackage{amssymb}
\usepackage{amsrefs}
\usepackage{bbold}
\usepackage{a4wide}
\usepackage{mathtools}
\usepackage{enumitem}

\newtheorem{theorem}{Theorem}[section]

\newtheorem{lemma}[theorem]{Lemma}
\newtheorem{definition}[theorem]{Definition}
\newtheorem{corollary}[theorem]{Corollary}

\begin{document}

\setlength{\parindent}{0mm}

\title{Two-way Unclonable Encryption with a vulnerable sender}

\author{Daan Leermakers, Boris \v{S}kori\'{c}\\
{\small Eindhoven University of Technology}
}

\date{}

\maketitle

\begin{abstract}
\noindent Unclonable Encryption, introduced by Gottesman in 2003 \cite{uncl},
is a quantum protocol that
guarantees the secrecy of a successfully transferred classical message 
even when all keys leak at a later time. 
We propose an Unclonable Encryption protocol with the additional property that the sender's key 
material is allowed to leak even in the case of 
an {\em unsuccessful} run. 
This extra feature makes it possible to achieve secure quantum encryption even when one of the parties is unable to protect its keys against after-protocol theft. Such an asymmetry occurs e.g.\;in case of server-client scenarios, 
where the client device is resource-constrained and/or located in a hostile environment.

Our protocol makes use of a bidirectional quantum channel in a manner similar to the
two-way protocol LM05 \cite{LM05}. 
Bob sends random qubit states to Alice. 
Alice flips the states in a way that depends on the message and a shared key, and sends the resulting states back to Bob.
Bob recovers Alice's message by measuring the flips.
We prove that our protocol satisfies the definition of unclonable encryption
and additionally that the message remains secure even if all of Alice's keys leak after the protocol.
Furthermore, we show that some of the key material can be safely re-used.
Our security proof is formulated in terms of diamond norms, which makes it composable,
and allows for noisy quantum channels.
We work out the details only for the asymptotics in the limit of long messages.

As a side result we construct a two-way QKD scheme with a high key rate.
We show that its key rate is higher than 
the rate of the two-way QKD scheme LM05 proven in
 \cite{BLMR2013} for the case of independent channel noise.
\end{abstract}

\section{Introduction}

\subsection{Motivation}
\label{sec:motivation}

Quantum cryptography uses the properties of quantum physics to achieve security feats which are impossible with only classical physics.
The best known example is Quantum Key Distribution (QKD), by which Alice and Bob create a shared secret key.
The QKD key is typically used for classical One-Time Pad (OTP) encryption.
Afterwards the key should never leak, for otherwise Eve can recover the plaintext.

Motivated by the difficulty of (i) fully wiping data from nonvolatile memory and 
(ii) keeping secrets forever, D.\,Gottesman in 2003 introduced Unclonable Encryption (UE) \cite{uncl}. 
In UE, Alice directly encodes the message into the quantum states. 
Bob responds with a feedback bit {\it accepting} or {\it rejecting} the communication depending on whether he detected a disturbance.
UE provides information-theoretic security even when all key material becomes public after an {\it Accept}.
This relieves Alice and Bob of the burden of eternal confidential storage or perfect deletion. 
In the {\it Reject} case, however, the burden is still there. 

Our aim in this paper is to allow for even more leakage than Gottesman's UE.
In particular, we want a scheme that allows one party to leak all its keys even in the {\it Reject} case.
Here it is important to remark on a fundamental impossibility.
The leaking party cannot be the {\em receiver} of the message (Bob).
Bob's keys are by definition sufficient for decrypting the cipherstate;
if the {\it Reject} was caused by Eve intercepting the complete cipherstate, and she gets Bob's keys,
then she can decrypt.  
Hence, the most we can aim for
is a scheme that allows leakage at Alice's side.\footnote{
Of course it is then allowed for Bob to leak keys that Alice too possesses.
}
Such a thing is possible only if Alice's keys are not the same as Bob's.

We achieve the asymmetry between Alice and Bob 
by using a quantum channel in two directions, in a way similar to
Quantum Secure Direct Communication (QSDC) and two-way QKD schemes \cite{LM05,pingpong,BLMR2013}.
Bob creates qubit states that are not fully known to Alice and 
`bounces' them off Alice, who modifies them in a message-dependent way.
Alice does not possess the secrets needed to read out the qubit states.

By relieving Alice of the burden of protecting or wiping her keys after a protocol run,
we reduce the number of ways in which the security of an UE scheme
can be compromised by an attacker.
A full break of the sender's device may now be allowed regardless of the Accept/Reject outcome.
We will call this the {\it Vulnerable-Sender} (VS) property, and refer to our scheme as a
VSUE scheme.
VSUE is useful especially in scenarios where the sender is a resource-constrained device, or located in
a hostile environment.

An additional motivation for allowing more leakage than UE 
is that 
Alice and Bob's shared keys may have been derived via some mechanism that does not have the UE property.
Then the keys that are shared between them are more vulnerable than keys
that exist only at Alice's or Bob's side.
In scenarios where Alice holds shared keys only, the situation is equivalent to 
the Vulnerable Sender setting sketched above.

\subsection{Related work}

\underline{Unclonable Encryption and Quantum Key Recycling}.
Unclonable Encryption was introduced by Gottesman in 2003 \cite{uncl} to extend the no-cloning property of quantum states to classical messages. By encoding a classical message into a quantum state directly, the inability of an attacker to clone the quantum state will guarantee the confidentiality of the message even if all classical information associated to the communication leaks afterward.\footnote{
An alternative definition of unclonable encryption exists \cite{BroadbentLord},
with two collaborating parties who attempt to both recover the plaintext.
We will not use this definition.
} 
Gottesman introduced a qubit-based prepare-and-measure protocol that achieves UE but uses up key material.

Methods for re-using key material have been developed under the term Quantum Key Recycling (QKR) 
\cite{BBB82,FehrSalvail2017,DBPS2014,QKR_noise,SilentBob}.
The main advantage of QKR compared to QKD is reduced round complexity.
The combination of QKR and UE was studied in \cite{KRUE}.
It was shown that key re-use and UE can be achieved simultaneously in a protocol 
similar to the first ever QKR scheme \cite{BBB82} and to the first qubit-based scheme
that has been proven secure \cite{FehrSalvail2017}.

\underline{Two-way quantum protocols}.
The back-and-forth use of a quantum channel has been studied mainly
because it enables {\em passive} correction of polarisation drift in fibers.\footnote{
Another motivation is a reduction of the number of qubits that have to be discarded because
of basis mismatch between Alice and Bob in typical QKD.
However, the gain is minor given that there are highly efficient biased-basis QKD schemes \cite{LoChauArdehali2004}
that need to discard only $\cO(\log n)$ qubits.
}
Several schemes have been proposed, 
most notably the ping-pong protocol \cite{pingpong} and LM05~\cite{LM05}. 
Ping-pong uses entanglement and a two-qubit measurement at Bob's side.
LM05 achieves the same communication but with single-qubit operations. 
In LM05,
Bob sends qubits to Alice in BB84 states ($x$-basis or $z$-basis on the Bloch sphere). 
Alice may choose to flip the state independent of the encoding basis by applying the Pauli operation $\qs_y$. 
Bob measures in the correct basis to see if a flip occurred.
This creates the communication channel from Alice to Bob. The same channel will be used in our two-way protocol.

Variants of ping-pong and LM05 have been used to construct QKD protocols. 
A QKD version of LM05 was proven secure against general attacks in \cite{HFMC2011}. 
A proof technique based on an entropic uncertainty relation improved the communication rate \cite{BLMR2013}.

\subsection{Contributions and outline}

We propose and study a two-way protocol that achieves unclonable encryption with improved {\it Reject} behavior. 
The sender's key material is allowed to leak independent of the Accept/Reject outcome.
\begin{itemize}[leftmargin=4mm,itemsep=0mm]
\item 
We introduce a security notion for quantum protocols that we call Vulnerable-Sender Unclonable Encryption (VSUE). 
Ordinary UE guarantees that all keys may be leaked after an Accept without jeopardising the message.
VSUE in addition guarantees that all Alice's keys are allowed to leak even upon {\em Reject}.
We formulate VSUE in terms of trace distance.

\item 
We introduce a protocol for the Vulnerable Sender setting.
It makes two-way use of the quantum channel, to make sure that 
Alice does not have to know all of Bob's keys.
We prove that our protocol satisfies VSUE and furthermore allows for partial key re-use.

Our proof is based on bounding diamond norms and hence it is composable.
The proof consists of the following steps.
(i)
First we construct an EPR version of the protocol that is equivalent to the original.
Eve produces noisy EPR pairs.
Per raw bit that is communicated from Alice to Bob, {\em two} EPR pairs are consumed. 
(ii)
We use the Postselection technique \cite{CKR2009} to relate security against general attacks to security against {\em collective} attacks
(attacks where Eve performs the same action in every position).
Here it is taken into account that the basic `unit' under attack consists of two EPR pairs.
(iii)
We apply noise symmetrisation with random Pauli operators \cite{RGK2005,RennerThesis} to Alice and Bob's two-EPR-pair state.
This results in a four-qubit mixed state with only two degrees of freedom, namely the bit error probabilities
in the individual EPR pairs.
In this step we need Alice and Bob's monitoring of the quantum channel to take place in at least three bases,
just as in the case of six-state QKD.
(iv)
We determine the output state of the protocol, comprising Eve's quantum system and all the relevant classical variables,
and we bound its trace distance to the `ideal' state in which Eve learns nothing important.
\item
The scarce resource in any quantum protocol are the qubits.
As a figure of merit we use the {\em rate}, which is defined as the useful-message length  divided
by the number of qubits that are expended in order to send the message.
Asymptotically our scheme achieves rate 
$(1-2J)(1-J)/(1-J+h(2\qb-2\qb^2))$, where $\qb$ is the tolerable bit error rate on the quantum channel,
$h$ is the binary entropy function and $J$ stands for 
$-(1-\fr32\qb)\log(1-\fr32\qb)-3\cdot\fr\qb2\log\fr\qb2$.
\item 
We present a side result of our work.
From our protocol we construct a two-way QKD variant.
We show that it has a slightly higher key rate than 
\cite{BLMR2013} in the case of independent channel noise.
\end{itemize}

We start by introducing notation, terminology and some useful previous results in the preliminaries (Section \ref{sec:prelims}).
In Section \ref{sec:secdef} we introduce the security definitions of Vulnerable-Sender Unclonable Encryption and key re-use.  
In Section \ref{sec:twoWayUE} we present our two-way protocol, and in Section \ref{sec:equivalent}
we introduce a modified but equivalent version for which the security can be more easily proven.
In Section \ref{sec:stateEve} we derive Eve's state after Alice and Bob have run the protocol. 
The security proof is presented in Section \ref{sec:secProof}. 
Our two-way QKD side result is discussed in Section \ref{sec:QKD}.

\section{Preliminaries}
\label{sec:prelims}

\subsection{Notation and terminology}
\label{sec:notation}

Classical Random Variables are denoted with capital letters, and their realisations
with lowercase letters. 
The expectation with respect to $X$ is denoted as 
$\EE_x f(x)=\sum_{x\in\cX}\pr[X=x]f(x)$.
For the $\ell$ most significant bits of the string $s$ we write $s[:$$\ell]$. 
The Hamming weight of a string $s$ is denoted as $|s|$.
The notation `$\log$' stands for the logarithm with base~2.
The notation $h$ stands for the binary entropy function $h(p)=p\log\fr1p+(1-p)\log\fr1{1-p}$.
Sometimes we write $h(p_1,\ldots,p_k)$ meaning $\sum_{i=1}^k p_i\log\fr1{p_i}$.
Bitwise XOR of binary strings $x,y$ is written as $x\oplus y$,
and bitwise logical {\it and} as $x \land y$.

The Kronecker delta is denoted as $\qd_{ab}$.
We will speak about
`the bit error rate $\qb$ of a quantum channel'.
This is defined as the probability that a classical bit $x$, sent by Alice embedded in a qubit,
arrives at Bob's side as the flipped value $\bar x$. 
The combined error rate of two subsequent channels is written as $\qb \star \qg = \qb(1-\qg) + (1-\qb)\qg$.
We will be working with linear error-correcting codes.
Asymptotically for large $n$, the size of the syndrome needed to correct
an error rate $\qb$ in a string of length $n$ is given by  $n h(\qb)$
(see e.g.\,\cite{BKB2004}).

For quantum states we use Dirac notation.
A qubit with classical bit $x$ encoded in basis $b$ is written as $\ket{\qj^b_x}$. 
The set of bases is~$\cB$.
In case of BB84 states we have $\cB=\{z,x\}$; in case of 6-state encoding
$\cB=\{z,x,y\}$. When performing computations with $b \in \cB$ we write $\{0,1,2\}$ instead of $\{z,x,y\}$.
The bit $a$ encoded in the z-basis is sometimes written as $\ket{a}$.
The notation $\ket{a}_x$ and $\ket{a}_y$ is used for the x- and y-basis respectively.
The Bell states are: $\ket{\Phi_{00}} = \frac{\ket{00}+\ket{11}}{\sqrt{2}}$, 
$\ket{\Phi_{uv}}=(\one \otimes \qs_x^u\qs_z^v) \ket{\Phi_{00}}=\frac{\ket{0,u}+(-1)^v\ket{1,\bar u}}{\sqrt2}$. 
The notation `tr' stands for trace.
Let $A$ have eigenvalues~$\ql_i$. 
The 1-norm of $A$ is written as $\|A\|_1=\tr\sqrt{A\dagg A}=\sum_i|\ql_i|$. 
The trace norm is $\| A\|_{\rm tr}=\fr12 \|A\|_1$.

We write $\cS(\cH)$ to denote the space of density matrices on Hilbert space $\cH$, 
i.e.\,positive semi-definite operators acting on $\cH$. 
States with non-italic label `A', `B' and `E' indicate the subsystem of Alice/Bob/Eve.
Consider classical variables $X,Y$ and a quantum system under Eve's control that depends on $X$ and~$Y$. 
The combined classical-quantum state is $\qr^{XY \rm E}=\EE_{xy} \ket{xy}\bra{xy} \otimes \qr^{\rm E}_{xy}$. 
The state of a sub-system is obtained by tracing out all the other subspaces, 
e.g. $\qr^{ Y \rm E}={\rm tr}_X \qr^{XY\rm E}=\EE_y \ket y\bra y\otimes\qr^{\rm E}_y$, with $\qr^{\rm E}_y=\EE_{x|y}\qr^{\rm E}_{xy}$.
The fully mixed state on $\cH_A$ is denoted as~$\chi^A$.

Any quantum channel can be described by a completely positive trace-preserving (CPTP) map 
$\cE: {\cS({\cH_{\rm A}})} \rightarrow {\cS(\cH_{\rm B})}$ that transforms a mixed state $\rho^{\rm A}$ 
to $\rho^{\rm B}$: $\cE(\rho^{\rm A}) = \rho^{\rm B}$.
For a map $\cE: \cS(\cH_{\rm A}) \rightarrow \cS(\cH_{\rm B})$, the notation $\cE(\rho^{\rm AC})$ stands for 
$(\cE \otimes \one_C) (\rho^{\rm AC})$, 
i.e.~$\cE$ acts only on the ‘A’ subsystem. 
The diamond norm of $\cE$ is defined as $\| \cE \|_\diamond =  \sup_{\rho^{\rm AC} \in \cS( \cH_{\rm AC})} \| \cE(\rho^{\rm AC})\|_{\rm tr}$ 
with ${\cH_{\rm C}}$ an auxiliary system that can be considered to be of the same dimension as $\cH_{\rm A}$. 
The diamond norm $\|\cE-\cE'\|_\diamond$ can be used to
bound the probability of distinguishing two CPTP maps $\cE$ and $\cE'$ given that the process is observed once. 
The maximum probability of a correct guess is $\frac12 + \frac14 \| \cE - \cE' \|_\diamond$. 
In quantum cryptography, one proof technique considers Alice and Bob performing actions on noisy EPR pairs.
These actions are described by a CPTP map $\cE$ acting on the input EPR states and outputting classical 
outputs for Alice and Bob, and correlated quantum side information for Eve.
The security of such a protocol is quantified by the diamond norm between the actual map 
$\cE$ and an idealised map $\cF$ which produces perfectly behaving outputs (e.g. perfectly secret QKD keys). 
When $\| \cE - \cF \|_\diamond \leq \qe$ 
we can consider $\cE$ to behave ideally except with probability $\qe$; 
this security metric is {\em composable} with other (sub-)protocols \cite{TL2017}.

We define the rate of a quantum communication protocol as the number of message bits communicated per sent qubit.
A protocol with a secret of length $\ell$ and security parameter $\qe$ which is dominated by a term $2^{\ell} \qe'$ has an improved security parameter when the secret size $\ell$ is decreased. We refer to the shortening of $\ell$ as privacy amplification. When the secret is the output of a pairwise independent hash function, this reduces to the definition of privacy amplification in \cite{RennerThesis}.

An information-theoretically secure MAC function can be constructed using pairwise independent hash functions and a single-use shared key \cite{WegmanCarter1981}. 
The probability of forging a tag is equal to the probability of randomly guessing the key or the tag. 
For a key and tag of the same size $\ql$ this probability is~$2^{-\ql}$.

\subsection{Pairwise independent hashing with easy inversion}
\label{sec:invertible_hash}

A family of hash functions $H=\{h:\cX\to\cT  \}$ is called pairwise independent 
(a.k.a.\,2--independent or strongly universal)
\cite{WegmanCarter1981} 
if for all distinct pairs $x,x'\in\cX$
and all pairs $y,y'\in\cT$ it holds that
$\pr_{h\in H}[h(x)=y \wedge  h(x')=y']=|\cT|^{-2}$.
Here the probability is over random~$h\in H$.
We will be using a family of invertible functions $F: \bits^\nu \to \bits^\nu$ that has the collision properties of a pairwise independent hash function.

An easy way to construct such a family is to use multiplication in $GF(2^\nu)$.
Let $a,b \in GF(2^\nu)$ be randomly chosen, and $u=(a,b)$. 
We define $F_u(x) = a \cdot x + b$, where the operations are in $GF(2^\nu)$. 
A pairwise independent family of hash functions $\qF$ from  $\bits^\nu$ to $\bits^\ell$, with $\ell \leq \nu$, is implemented by taking the $\ell$ 
most significant bits of $F_{u}(x)$.\footnote{
The proof is straightforward. Write $F_u(x)=c\|r$, with $c\in\bits^\ell$.
Let $x'\neq x$.
Then $\pr_u[\qF_u(x)=c \wedge \qF_u(x')=c']$
$=\sum_{r,r'\in\bits^{\nu-\ell}} \pr_u[F_u(x)=c\|r \;\wedge\; F_u(x')=c'\|r']$.
By the pairwise independence of $F$ this gives 
$\sum_{r,r'\in\bits^{\nu-\ell}} 2^{-2\nu}=2^{-2\ell}$.
}  
We denote this as
\be
	\qF_{u}(x) = F_{u}(x)[:\!\ell].
\ee
The inverse operation is as follows. Given $c\in\bits^\ell$, generate random $r\in\bits^{\nu-\ell}$ and output $F_{u}^{\rm inv}(c||r)$.
It obviously holds that $\qF_{u}( F_{u}^{\rm inv}(c||r) )=c$.
Note that we don't rely on the properties of the inverse operation but only on the properties of $\Phi_{u}$.
Computing an inverse in $GF(2^\nu)$ costs $O(\nu \log^2 \nu)$ operations \cite{Moenck73}.

\subsection{Smooth R\'{e}nyi entropies}
\label{sec:smooth}

Let $\qr$ be a mixed state.
The $\qe$-smooth R\'{e}nyi entropy of order $\qa$ is defined as \cite{RK2005} 
\be
	\mbox{For }\qa\in(0,1)\cup(1,\infty): \quad\quad
	S_\qa^\qe(\qr) \isdef \frac1{1-\qa}\log \min_{\bar\qr:\, \|\bar\qr-\qr\|_1\leq\qe} \tr \bar\qr^\qa,
\ee
where the density operator $\bar\qr$ may be sub-normalised.
Furthermore $S_0^\qe(\qr)=\lim_{\qa\to0}S_\qa^\qe(\qr)$
and $S_\infty^\qe(\qr)=\lim_{\qa\to\infty}S_\qa^\qe(\qr)$
It has been shown that the smooth R\'{e}nyi entropy of factor states $\qr^{\otimes n}$
asymptotically approaches the von Neumann entropy $S(\rho) = -\tr\rho \log \rho$. 
In particular, asymptotically for large $n$ and fixed $\qe$ it holds that \cite{RennerThesis}
\bea
	S_2^\qe(\qr^{\otimes n})&\to& nS(\qr)
	\\
	S_0^\qe(\qr^{\otimes n}) &\to& nS(\qr).
\eea

\subsection{Post-selection}
\label{sec:post-selection}

It has been shown \cite{CKR2009} that
any protocol that is invariant under permutation of its inputs and secure against {\em collective} attacks 
(same attack applied to each qubit individually), 
is also secure against {\em general} attacks, at the cost of some extra privacy amplification. 
Let $\cE$ be a protocol that acts on $\cS(\cH_{\rm AB}^{\otimes n})$
and let $\cF$ be the `ideal' version of $\cE$, i.e.\;having perfect security.
If for all input permutations $\pi$ there exists a map $\cK_\pi$ on the output such that $\cE \circ \pi = \cK_\pi \circ \cE$, then
\bea
	\| \cE -\cF \|_\diamond &\leq& (n+1)^{d^2-1} \max_{\qs \in \cS(\cH_{\rm ABE})} \Big\| (\cE - \cF)  ( \qs^{\otimes n})\Big\|_1  
\label{eq:post-selection}
\eea
where $d$ is the dimension of the $\cH_{\rm AB}$ space. ($d=4$ for qubits).
We will be working with two EPR pairs, i.e.\;$d=16$.
The product form $\qs^{\otimes n}$ greatly simplifies the security analysis:
now it suffices to prove security against {\em collective} attacks,
and to pay a price $2(d^2-1)\log(n+1)$ in the amount of privacy amplification, which is negligible compared to $n$
for $n\gg 1$.

\section{Security Notions}
\label{sec:secdef}
We are concerned with protocols in which a classically authenticated classical message $m$ is communicated over a quantum channel. 
We briefly introduce the attacker model and the security definitions.

\subsection{Attacker Model}
\label{sec:attacker}

We distinguish between two kinds of secret data.
(i) 
Secrets which, upon being generated or received, are used `on the fly'
and are immediately discarded. 
They are needed for a short time, and in volatile memory only; thus they are easy to delete entirely.
We will refer to these secrets as {\em volatile}.
(ii)
Secrets which are kept, at some point in time, in nonvolatile memory.
We take a very conservative approach and assume that the act of waiting for a communication to arrive
causes such a long delay that data gets stored in nonvolatile memory.
We will refer to these secrets as {\em nonvolatile keys} or simply {\em keys}.

We assume that volatile secrets will never leak to the adversary.
Nonvolatile keys are harder to protect, however, and we adopt a particular model that describes
whose keys can leak and when.
Here it is important to note that we will be considering Quantum Key Recycling. 
In QKR, keys are re-used after an Accept and refreshed after a Reject.
During a certain time window, which depends on the context,
Alice and Bob are confident that their keys have not yet leaked, and are re-using keys upon Accept.
(In \cite{KRUE} this period was defined as $N$ protocol runs, with $N$ some constant.)
The assumption is that there is indeed no leakage during this time window.\footnote{
Without such an assumption there is no meaningful way of doing QKR in a setting that requires UE.
}
Furthermore, upon a {\em Reject} Bob (but not Alice) is able to permanently destroy or indefinitely protect his keys that were
associated with the rejected run.
Apart from those, all nonvolatile keys are assumed to leak after the time window.

The rest of the attacker model consists of the standard assumptions: 
No information, other than specified above, leaks from the labs of Alice and Bob; there are no side-channels; 
Eve has unlimited quantum storage and computing resources; 
all noise on the quantum channel is considered to be caused by Eve.

We do not consider automatic polarisation drift correction by double use of the channel.
The noise in the two uses of the quantum channel is assumed to be entirely independent.

We will not explicitly write out the message authentication steps.
Instead we give Alice and Bob access to an authenticated classical channel. 
It is understood that every use of this classical channels adds an error term 
$2^{-\nu}$ to the final diamond distance (with $\nu$ the security parameter of the MAC), 
and that Alice and Bob need shared MAC keys.

\subsection{Security definitions}
\label{sec:securitydefs}

We first present the semantics for quantum encryption protocols.

\begin{definition}
\label{def:QE}
A Quantum Encryption scheme {\sf QE} with message space $\cM$, key space $\cK$ and Hilbert space $\cH$
consists of the following components:
\begin{enumerate}[leftmargin=5mm,itemsep=0mm]
\item
A key generation function 
{\sf QE.Gen}$:1^\ql\to \cK$, where $\ql$ is the security parameter.
\item
A CPTP map
{\sf QE.Encr}$:\cM\times \cK\times \cS(\cH) \to \cM\times\cK\times\cS(\cH)$ 
that takes as input a message $m\in\cM$ and a key $k\in\cK$, acts on an initial state
$\pi^0\in\cS(\cH)$, and outputs a cipherstate $\pi_{mk}\in\cS(\cH)$ which may or may not
contain a classical part.
(The $m$ and $k$ are not modified.)
We write
${\sf QE.Encr}\Big(\ket{mk}\bra{mk}\otimes\pi^0 \Big)=
\ket{mk}\bra{mk} \otimes  \pi_{mk}$.
\item
A measurement {\sf QE.Decr}$:\cK\times\cS(\cH) \to \cK\times\cS(\cM\cup\{\bot\})\times\bits$ 
that takes as input a key $k\in\cK$ and a cipherstate, and outputs 
(i)
a message $m'\in\cM$ or the error message~$\bot$;
(ii)
a flag $\qo$ which is set to $\qo=0$ ({\tt reject}) if $m'=\bot$, 
and to $\qo=1$ ({\tt accept}) if $m'\in\cM$.
(The $k$ is not modified.)
\end{enumerate}
\end{definition}

The {\sf QE.Gen} is a joint procedure run by Alice and Bob.
Typically the {\sf QE.Encr} is done by Alice, who sends the cipherstate $\pi_{mk}$ to Bob.
Eve couples some ancilla state of her own to $\pi_{mk}$ and forwards the modified state
$\pi_{mk}'$ to Bob.
This adversarial action is represented as a CPTP map~$\cA$ acting on the cipherstate and Eve's ancilla,
which has starting state $\ket e\bra e$.
Then {\sf QE.Decr} is done by Bob, and he sends the feedback $\qo$ to Alice.

\underline{Correctness}.
Correctness means that $m'=m$ with near certainty when $\qo=1$ occurs. 
Furthermore, in the next run of the protocol Bob's keys should equal Alice's keys with overwhelming probability.
For protocols like ours (Section~\ref{sec:twoWayUE}), where the message is authenticated with a classical MAC 
and the re-used keys are unaltered, 
correctness is guaranteed except with probability $2^{-\nu}$, the probability of forging the MAC.

We use the term `output state' or `final state' for the mixed state
that describes all the important classical variables as well as Eve's ancilla, 
after {\sf QE.Decr}.
Since Eve's ancilla may contain a classical part (Def.\ref{def:QE} allows for a classical part of the ciphertext)
we split Eve's ancilla into a purely quantum part `E' and a classical {\em transcript}~$T$.
The splitup will make it easier to formulate an EPR version of a protocol. 
We write the output state as $\qr^{MKM' \qO T\rm E}$.

\begin{definition}
\label{def:ENC}
Let {\sf QE} be a quantum encryption scheme according to Def.\,\ref{def:QE} with output state
$\qr^{MKM'\qO T\rm E}$ 
as described above.
{\sf QE} is called {\bf $\qe$-encrypting ($\qe$-ENC)} if the output state satisfies
\be
	 \Big\|\qr^{M\qO T\rm E}-\qr^M\otimes\qr^{\qO T\rm E}\Big\|_1\leq\qe
\label{ENCdef}
\ee
for all adversarial actions $\cA$ and all distributions of~$M$.
\end{definition}
(The $\qe$ is referred to as the {\em error}).
Note that $M'$ has been traced out since Eve does not care about~$M'$.
Def.\,\ref{def:ENC} states that the message $M$ is independent of Eve's information
$\qO T\rm E$ except with probability~$\qe$.

Note that Def.\,\ref{def:ENC} demands that (\ref{ENCdef}) holds for all distributions of $M$;
hence it implies other security definitions used in the literature which
work with `$\forall_m$' conditions.

The output state of {\sf QE} can be written as
$\qr^{MKM' \qO T\rm E}=\qr^{MKM' T\rm E}_{\rm accept}+\qr^{MKM' T\rm E}_{\rm reject}$,
where the two states in the right hand side are sub-normalised, with
$\tr \qr^{MKM' T\rm E}_{\rm accept}=\pr[\qO=1]$ and
$\tr \qr^{MKM' T\rm E}_{\rm reject}=\pr[\qO=0]$.
{\em Unclonable Encryption} is a requirement on the {\em accept} part. 

\begin{definition}
\label{def:UE}
Let {\sf QE} be a quantum encryption scheme according to Def.\,\ref{def:QE} with output state
$\qr^{MKM'\qO T\rm E}$.
{\sf QE} is called an {\bf $\qe$-Unclonable ($\qe$-UE)} if
\be
	\Big\|\qr^{MK T\rm E}_{\tt accept}-\qr^M\otimes\qr^{K T\rm E}_{\tt accept}\Big\|_1\leq\qe
\label{UEdef}
\ee
for all adversarial actions $\cA$ and all distributions of~$M$.
\end{definition}

Def.\,\ref{def:UE} states that either the accept probability is very low due to Eve's interference, 
or else $M$ is almost independent of Eve's information even if she learns $K$ after an Accept.

{\bf Remark}.
Gottesman's definition of unclonable encryption \cite{uncl} demands that 
$\| \qr^{\rm F}_{\rm accept}(m,k)-\qr^{\rm F}_{\rm accept}(m',k) \|_1\leq\qe$ for a fraction $\geq 1-\qe$ of 
keys $k\in\cK$ and for all pairs $m,m'\in\cM$, $m'\neq m$;
this property is implied by our Def.\,\ref{def:UE}.\footnote{
(i) For the specific values $m,m'$ the same reasoning applies as with Def.\,\ref{def:ENC}.
(ii) Our definition works with an average over $k$, and hence the desired $\|\cdots\|_1\leq\qe$ property
may fail to hold for a fraction $\qe$ of all values $k\in\cK$. This is the same fraction as in
Gottesman's definition.
}

In order to write down the definition of VSUE we need to distinguish between
{\em shared} keys `$S$', which Alice and Bob both have,
 and keys that are uniquely in Bob's possession,~`$P$'.
(Here we assume that Alice possesses shared keys only.)

\begin{definition}
\label{def:VSUE}
Let {\sf QE} be a quantum encryption scheme according to Def.\,\ref{def:QE} with output state
$\qr^{MKM'\qO T\rm E}$, where $K=(S,P)$, with $S$ shared keys.
{\sf QE} is called {\bf Vulnerable-Sender Unclonable} with error $\qe$ {\bf ($\qe$-VSUE)} if
it is $\qe$-UE and also satisfies
\be
	\Big\|  \qr^{MS T\rm E}_{\tt reject} 
	-\qr^M\otimes\qr^{S T\rm E}_{\tt reject}
	\Big\|_1 \leq\qe
\ee
for all adversarial actions $\cA$ and all distributions of~$M$.
\end{definition}

Def.\,\ref{def:VSUE} states that, on top of the UE property,
$M$ is secure also if the shared keys leak after a {\em Reject}.
Since we will restrict ourselves to schemes where Alice possesses shared keys only,
this leak represents a full compromise of Alice's keys.
Note that $\qe$-VSUE implies $\qe$-ENC and $\qe$-UE.

Finally, we are also interested in (partial) key re-use. 
The $P$, the one-sided keys, can be randomly re-generated
`for free' after each protocol run, because no communication is needed for such a refresh.
Hence the question of key re-use is relevant only for the shared keys~$S$.

\begin{definition}
\label{def:KRinVSUE}
Let {\sf QE} be a quantum encryption scheme according to Def.\,\ref{def:QE} with output state
$\qr^{MKM'\qO T\rm E}$, where $K=(S_{\rm re},S_{\rm once},P)$, with $S_{\rm re},S_{\rm once}$ shared keys.
{\sf QE} is called {\bf $\qe$-Key-Reusing ($\qe$-KR)} if
\be
	\Big\| \qr^{MS_{\rm re}S_{\rm once}P\qO T\rm E}  -  \qr^{S_{\rm re}}\otimes\qr^{MS_{\rm once}P\qO T\rm E}
	\Big\|_1 \leq \qe
\label{KRcondition}
\ee
for all adversarial actions $\cA$ and all distributions of~$M$.
\end{definition}
\underline{Remark 1}.
In Def.\,\ref{def:KRinVSUE} Eve potentially gets access to part of the key material.
Thus Def.\,\ref{def:KRinVSUE} imposes requirements that are much more demanding than 
other definitions of key recycling or key re-use.
\newline
\underline{Remark 2}.
In order for a scheme to satisfy VSUE, Eve must not have access to~$P$.
The reject-case output state of a VSUE scheme is of the form 
$\qr^{MS_{\rm re}S_{\rm once}P\qO T\rm E}_{\rm reject}=\qr^P\otimes\qr^{MS_{\rm re}S_{\rm once}\qO T\rm E}_{\rm reject}$.

\section{The protocol}
\label{sec:twoWayUE}

\subsection{Protocol intuition}
\label{sec:intuition}

As mentioned in Section~\ref{sec:motivation},
we create the knowledge asymmetry between Alice and Bob by making use of a two-way protocol,
where Bob `bounces' random qubit states off Alice, which states Alice is then able to flip without knowing them. 
A message from Alice to Bob is encoded in these flips.
Only Bob knows how to interpret the states coming back from Alice.

Channel monitoring has to occur on both the Bob-to-Alice channel (`Channel~1') and 
the Alice-to-Bob channel (`Channel~2').
This monitoring is done by sending, interspersed in between the ordinary states, {\em test qubits}
whose states are known to the receiver; the locations and the states are part of the a priori shared key material. 
In contrast to the data carrying qubits, which are in BB84 states (on the $xz$-circle on the Bloch sphere), 
we let the test bit states cover both dimensions of the Bloch sphere
(by using the same six states as six-state QKD \cite{sixStateOriginal,Bruss1998}).
The advantage of this larger test space is that the noise symmetrisation technique of Renner et al. \cite{RGK2005,RennerThesis}
then yields a higher rate than BB84 test states would.

Alice is willing to send meaningful data to Bob only if Channel~1 is sufficiently noiseless.
However, she already has to send data before she can assess the noise level.
The solution is to first send a random string and later decide to use that string as an encryption mask
for useful data.

Error correction and privacy amplification are done in a fairly straightforward way,
except for one technicality:
The privacy amplification must be computable in both directions.
We implement this with the hash functions discussed in Section~\ref{sec:invertible_hash}.

The shared secret hash seed can be re-used as is the case in QKR schemes \cite{QKR_noise,SilentBob} 
where it plays a similar role.

\subsection{Preparation}
\label{sec:protocol}

Alice holds a classical message $m\in\bits^\ell$ which contains an authentication tag
created with a one-time MAC.
The security parameter of the MAC is a constant that we will ignore in our analysis since we 
are mainly interested in the asymptotics.

There are $n + \ql$ qubits sent back and forth, $n$ for communication and $\ql$ for channel monitoring.
$\ql=\cO(\log n)$.
Alice and Bob share key material 
$k = (u, k_{\rm syn}, k_{\rm test}, \cI_{\rm test}, b_{\rm test}^1,b_{\rm test}^2, \xi, \qh)$. 
Here $u\in \bits^{2n}$ is the random seed used for privacy amplification; 
the $k_{\rm syn} \in \bits^{n-\qk}, k_{\rm test}\in\bits^{\ql}$ are used as One-Time Pads protecting the syndrome 
and Alice's test measurement outcome respectively; 
$\cI_{\rm test}\subset\{1,\ldots,n+\ql\}$, with $|\cI_{\rm test}|=\ql$,
describes the positions of the test qubits; 
$b^{1,2}_{\rm test} \in \{0,1,2\}^\ql$ are the bases in which the test qubits are prepared, 
and $\xi,\qh \in \bits^\ql$ are the payloads of the test qubits in Channel 1 and 2 respectively. 
In addition, not named explicitly, Alice and Bob share three authentication keys, one for the plaintext message $m$ and
two for the classical channel.

Alice and Bob agree on a efficiently invertible paiwise independent hash $\Phi_u:\bits^n\to\bits^\ell$, see Section \ref{sec:invertible_hash}. 
They agree on an error correcting code with
syndrome function ${\tt Syn}$: $\bits^n \rightarrow \bits^{n-\qk}$ and decoding function $\tt SynDec$: $\bits^{n-\qk} \rightarrow \bits^n$. 
They agree on noise thresholds $\qb^*,\qg^*$ for quantum Channels 1 and 2 respectively.
They adopt the channel monitoring procedure shown below, which accepts only if the bit error rates
in the test states are sufficiently low and
independent of the bases.

\begin{definition}
\label{def:noisechecks}
Let $\xi'$ be the noisy version of the test string $\xi$ as received by Alice through Channel~1, and
likewise let $\qh'$ be the noisy version of $\qh$ received by Bob.
For $b\in\{0,1,2\}$ let $\xi(b)$ denote the part of $\xi$ that is encoded in basis~$b$,
and likewise for $\xi',\qh,\qh'$.
Let $\tri_1(b)=\xi(b)\oplus\xi'(b)$ and $\tri_2(b)=\qh(b)\oplus\qh'(b)$
be error vectors.
For $b,b'\in\{0,1,2\}$ let $\xi(b,b')$ denote $\xi$ restricted to those positions where
the encoding basis is $b$ in Channel~1 and $b'$ in Channel~2.
Let $d_1(b,b')=\xi(b,b')\oplus\xi'(b,b')$ and $d_2(b,b')=\qh(b,b')\oplus\qh'(b,b')$
be the corresponding error vectors.
Let $\qd$ be a small parameter.
The channel monitoring consists of two verifications,
\bea
	{\sf CheckA}( \tri_1 ) &=& \begin{cases}
	1 & {\rm if}\;\exists_{\qb\leq\qb^*}\forall_{b\in\{0,1,2\}} |\frac{|\tri_1(b)|}{\ql/3}-\qb|\leq\qd \qb
	\\
	0 & {\rm otherwise}.
	\end{cases}
	\\
	{\sf CheckB}(\tri_1,\tri_2,d_1,d_2) &=& \begin{cases}
	1 & {\rm if}\;\exists_{\qb\leq\qb^*,\qg\leq\qg^*}\forall_{b,b'\in\{0,1,2\}}\Big\{
	|\frac{|d_1(b,b')\wedge d_2(b,b')|}{\ql/9}-\qb\qg|\leq\qd\sqrt2 \qb\qg
	\\ & {\rm and}\; |\frac{|\tri_1(b)|}{\ql/3}-\qb|\leq\qd \qb
	\\ & {\rm and}\; |\frac{|\tri_2(b')|}{\ql/3}-\qg|\leq\qd \qg
	\Big\}
	\\ 0 & {\rm otherwise}.
	\end{cases}
\eea

\end{definition}

Note that asymptotically for $n\to\infty$ the parameter $\qd$ can go to zero as $\cO(1/\sqrt n)$.
Our motivation for including a separate test for each $(b,b')$ combination is that it imposes symmetry,
which simplifies the noise symmetrisation step (Section~\ref{sec:paulitrick}).

\subsection{Protocol steps}
\label{sec:steps}

\begin{figure}
\center
\includegraphics[width=.9\textwidth]{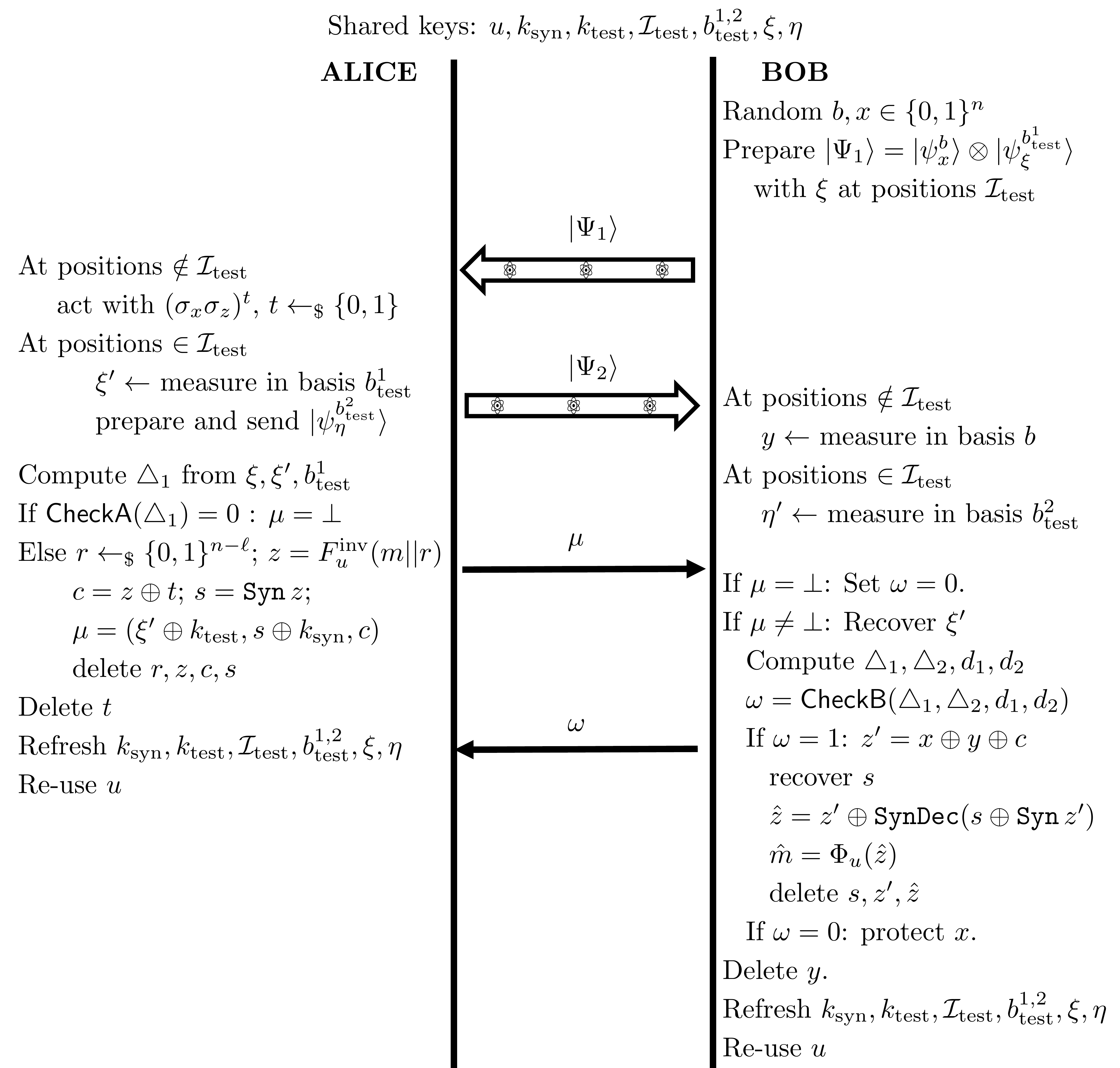}
\caption{\it 
Protocol steps.
}
\label{fig:protocol}
\end{figure}

Alice and Bob perform the following actions. (See Fig.\,\ref{fig:protocol}).\\

\underline{\bf Bob}: \\
Generate random $x,b \in \bits^{n}$. 
Prepare $n+\ql$ qubits. 
At locations $\cI_{\rm test}$ encode $\xi$ in basis $b^1_{\rm test}$. 
In the other positions encodes $x$ in basis $b$. 
Send the qubits to Alice. 
Store $x,b$ as private keys.

\underline{\bf Alice}: \\
At the $i$'th non-test position apply the Pauli operation $(\qs_x\qs_z)^{t_i}$ to the qubit,\footnote{
This operation has the effect that a BB84 state $\ket{\qj^b_x}$ is changed to $\ket{\qj^b_{x\oplus t_i}}$.
} 
where $t_i\in\bits$ is randomly generated on the fly, 
and send the resulting state back to Bob.
Remember the random bits to form a string $t\in\bits^n$.
In the test positions,
measure the qubits in basis $b^1_{\rm test}$, yielding measurement outcomes that together form a string
$\xi'\in\bits^\ql$.
In the $j$'th test position prepare state $\ket{\qj^{b^2_{\rm test}[j]}_{\qh[j]}}$
and send it.
Compute $\tri_1(0), \tri_1(1), \tri_1(2)$ from $\xi,\xi',b^1_{\rm test}$.
Perform {\sf CheckA}$(\tri_1)$. 
If the result is~$0$ then set $\mu=\bot$.
Otherwise 
\{take random $r \in \bits^{n-\ell}$;
Compute $z = F_u^{\rm inv}(m\|r)$,
$c=z\oplus t$
and
$s = {\tt Syn}\, z$;
set $\mu=(\xi' \oplus k_{\rm test}, s \oplus k_{\rm syn},c)$;
Delete $r,z,c,s$\}.
Over the authenticated classical channel send  $\mu$.
Delete $t$.

\underline{\bf Bob}: \\
At the non-test positions, measure the qubits in basis $b$, yielding $y\in\bits^n$. 
At the test positions, measure in basis $b^2_{\rm test}$, yielding $\qh'\in\bits^\ql$.
If Alice's classical message is $\bot$ then set $\qo=0$.
Otherwise \{Recover $\xi'$ from $\xi' \oplus k_{\rm test}$.
Compute $\tri_1,\tri_2,d_1,d_2$ from $\xi,\xi'\qh,\qh',b^1_{\rm test},b^2_{\rm test}$.
Perform $\qo=${\sf CheckB}$(\tri_1,\tri_2,d_1,d_2)$.
If $\qo=0$ take effort to protect~$x$.
If $\qo=1$ then \{
Compute $z' = x \oplus y \oplus c$.
Recover $s$ from $s \oplus k_{\rm syn}$.
Perform error correction as
$\hat z = z' \oplus {\tt SynDec}(s \oplus {\tt Syn}\, z')$.
Reconstruct the message $\hat m = \Phi_u(\hat z)$.
Delete $s,z',\hat z$.\} \\
Delete $y$.
Send $\qo$ over the authenticated channel.

\underline{\bf Key update}:\\
Alice and Bob refresh $k_{\rm syn}, k_{\rm test},\cI_{\rm test},b^{1,2}_{\rm test},\xi,\qh$.
They re-use~$u$.

\vskip5mm

Alice's {\em nonvolatile} keys are the shared keys.
Her {\em volatile} secrets are $t,r,z,c,s$.
Bob's nonvolatile keys are the shared keys and in addition $b,x$.
His volatile secrets are $s,z',\hat z,y$.
The deletion of volatile secrets from RAM memory is explicitly indicated in the protocol.

\clearpage

\section{Modified protocol for the security proof}
\label{sec:equivalent}

\subsection{List of modifications}

We will prove the security of a modified scheme which is based on EPR pairs.
Security of the modified scheme implies security of the original scheme.
In Sections \ref{sec:EPRbounce}--\ref{sec:paulitrick} we introduce protocol modifications
step by step. 
At each step we indicate why the newly obtained scheme is equivalent, security-wise,
to the previous.

The main modifications are
(i) EPR re-formulation of Bob's state preparation and Alice's state manipulation;
(ii) EPR re-formulation of the channel monitoring;
(iii) A random permutation of the EPR pairs, which makes the existing permutation symmetry explicitly visible. 
Permutation symmetry is needed in order to apply Post-selection (Section~\ref{sec:post-selection});
(iv) Noise symmetrisation using random Pauli operators.

Furthermore, the use of the one-time pads $k_{\rm test}$ and $k_{\rm syn}$
is replaced by a confidential channel, invisible to Eve, through which $\xi'$ and $s$ respectively
are transported.
The length of $k_{\rm test},k_{\rm syn}$ is still taken into account when we compute
communication efficiency.
Similarly, the existence of the shared keys $\xi,\qh$ is replaced by a confidential channel
over which to transport data of the same size.

Note that the modified scheme requires Alice and Bob to have quantum memory.
This does not affect the practical implementability of the original scheme,
since the EPR-based scheme serves as a proof-technical construct only.

In Section~\ref{sec:modified} we give the full details of the modified scheme.

\subsubsection{EPR variant of the `bounce'}
\label{sec:EPRbounce}

We replace Alice's operation $(\qs_x\qs_z)^t$ on the qubit state $\ket{\qj^b_x}$ (and Bob's recovery of $t$) by the following steps.
Let $\cA_0=\{\one,\qs_y\}$, $\cA_1=\{\qs_z,\qs_x\}$.
First Alice flips a coin $a\in\bits$ which decides whether she will pick $\cA_0$ or $\cA_1$.
Then she applies $\cA_a[t]$ to $\ket{\qj^b_x}$.\footnote{
I.e. $(a,t)=(0,0)$ corresponds to $\one$, $(0,1)$ to $\qs_y$, $(1,0)$ to $\qs_z$ and $(1,1)$ to $\qs_x$.
Alice's action can be seen as a Quantum One Time Pad encryption $\qs_x^t \qs_z^{a\oplus t}$.
}
She sends the resulting qubit state and $a$ to Bob.
Bob measures in basis $b$, yielding $y\in\bits$.
In case $a=0$, Bob recovers $t$ as $t=y\oplus x$ as before.
In case $a=1$, a flip in basis $b=0$ results from the $\qs_x$ while a flip in basis $b=1$ results from $\qs_z$;
hence $t=x\oplus y\oplus b$.
Overall $t=x\oplus y\oplus ab$.

Note that Eve learns nothing about $t$ by observing~$a$.
Obviously, if this protocol variant with the additional parameter $a$ is secure then 
the original protocol is secure.

Next we make one more change.
Let $\ket{\qF_{vw}}$ be the Bell states on the two-qubit space
as defined in Section~\ref{sec:notation}.
We replace the sending of a random bit $t\in\bits$ by the following steps.
Eve prepares two (noisy) EPR pairs, nominally in the state $\frac{\ket{00}+\ket{11}}{\sqrt2}$,
and sends one half to Alice, one half to Bob.
Alice measures her two qubits in the Bell basis $\ket{\qF_{vw}}$, yielding outcome $(v,w)\in\bits^2$.
She sets $t=v$, $a=v\oplus w$. She sends $a$.
Bob generates random $b\in\bits$ and measures both his qubits in basis $b$, yielding $x,y\in\bits$.
He recovers $t$ as $t=x\oplus y\oplus ab$.\footnote{
This procedure can be interpreted as entanglement swapping by Alice, so that Bob's qubits become entangled,
but with one difference: Bob receives only one classical bit ($a=v\oplus w$) instead of the two `key' bits $v,w$. 
Alternatively, it can be viewed as an incomplete teleport. First, Bob's measurement in basis $b$, yielding $x$,
is equivalent to sending $\ket{\qj^b_x}$. 
Alice's Bell measurement is a teleport, turning the state of Bob's second qubit into a random encryption
$\qs_x^v\qs_z^w\ket{\qj^b_x}$.
For a full teleport Alice would send $v$ and $w$.
}
Security of the EPR variant implies security of the protocol variant described above.
The same equivalence is used in \cite{BLMR2013}.

\subsubsection{EPR variant of the channel monitoring}
\label{sec:EPRmonitor}

The monitoring procedure of sending a state $\ket{\qj^b_\xi}$ (with shared $b\in\{0,1,2\},\xi\in\bits$)
to the other side, where it is measured in basis $b$, is now replaced by the following procedure.
The basis $b$ is still a shared secret.
Eve prepares two (noisy) EPR pairs, nominally in the state $\ket{\qF_{00}}$,
and sends one half to Alice, one half to Bob.
Alice measures her qubit in basis $b$, obtaining outcome $\xi^{\rm A}$.
Bob measures his qubit in basis $b$, obtaining outcome $\xi^{\rm B}$.
One party sends its measurement outcome to the other side, in plaintext.
There the noise bit $\tri$ is computed as $\tri=\xi^{\rm A}\oplus\xi^{\rm B}\oplus\qd_{b2}$.
(The contribution $\qd_{b2}$ comes from the fact that the $\ket{\qF_{00}}$ state\footnote{
If we had chosen the singlet state $\propto \ket{01}-\ket{10}$ there would be no asymmetry between the bases.
}
causes $\xi^{\rm B}=1-\xi^{\rm A}$ when $b=2$.)
This procedure is followed in all test positions $\cI_{\rm test}$
and allows for the computation of the strings $\tri_1(b),\tri_2(b),d_1(b,b'),d_2(b,b')$ as in the original protocol.

\subsubsection{Random permutation}
\label{sec:perminv}

One may argue that the original protocol (Section~\ref{sec:steps}) does not have symmetry under permutation 
of the locations $\{1,2,\cdots,n+\ql\}$ because the keys already exist at the beginning of the protocol.
For this reason we now include the key generation as part of the protocol.
(Obviously, generation of random keys is permutation invariant.)
This has no effect on the security analysis.
We let the key generation take place {\em after} the EPR pairs have been distributed;
in this way it becomes possible to formulate the whole protocol as a mapping that acts purely
on the qubits received by Alice and Bob.

To make the permutation invariance of the modified protocol explicitly visible, we introduce a permutation
that is applied to the EPR states before anything else is done.
Of course we have to justify why this addition has no impact on the security analysis.
A permutation of the EPR positions re-distributes the noise.
First, this has no impact on the statistics of the error counts
$|\tri_1(b)|$, $|\tri_2(b)|$, $|d_1(b,b')|$, $|d_2(b,b')|$ (since $\cI_{\rm test}$ is random and unknown to Eve) 
and hence the monitoring is not affected.
Second, the error correction of $z'$ is not sensitive to permutation of the bit flips.

Let $\cE_{\rm perm}$ denote the CPTP map describing the new protocol, which has the random permutation as its first step.
Permutation invariance is evident since we have $\cE_{\rm perm} \circ \pi = \cE_{\rm perm}$ for any permutation~$\pi$.
This allows us to use the post-selection Theorem  (Section~\ref{sec:post-selection}) and focus on {\em collective} attacks.
Crucially, this means that {\em the noise level caused by Eve is the same in the non-test locations as in the test locations},
and has to be such that the monitoring tests are passed.

\subsubsection{Noise symmetrisation with random Pauli operators}
\label{sec:paulitrick}

We apply the noise symmetrisation trick as introduced in \cite{RennerThesis}.
Alice and Bob publicly draw random strings $\qa_1,\qa_2\in \{0,1,2,3\}^{n+\ql}$.
Before they do any measurement, they each apply the Pauli\footnote{
Here $\qs_0$ stands for the $2\times 2$ identity matrix.
}
 operation 
$\qS(\qa_1)=\bigotimes_{i=1}^{n+\ql}\qs_{\qa_1[i]}$ 
on their own EPR qubits in Channel~1,
and $\qS(\qa_2)$ in Channel~2.
Then they forget $\qa_1,\qa_2$.
This results in a huge simplification of Alice and Bob's state:
only 15 global parameters are left to describe the whole state.

Of course we have to justify why adding this operation does not affect the correctness and the security of the protocol.
All measurements in the protocol (with outcomes $x,y,a,t$) are done in either the $x$, $y$ or $z$ basis.
For $b\in\{0,1,2\}$
the effect of a Pauli on a qubit state $\ket{\qj^b_x}$ is at most a bit flip to $\ket{\qj^b_{\bar x}}$. 
Thus, the effect on the outcomes $x,y,a,t$ is at most a number of bit flips;
but since Alice and Bob apply the same Paulis, the flips do not prevent Bob from reconstructing the correct~$t$.
Furthermore, Eve {\em knows} $\qa_1,\qa_2$, so for her the statistics of $x,y,a,t$
have not changed.

\subsection{The modified protocol}
\label{sec:modified}

\begin{figure}[!ht]
\center
\includegraphics[width=.9\textwidth]{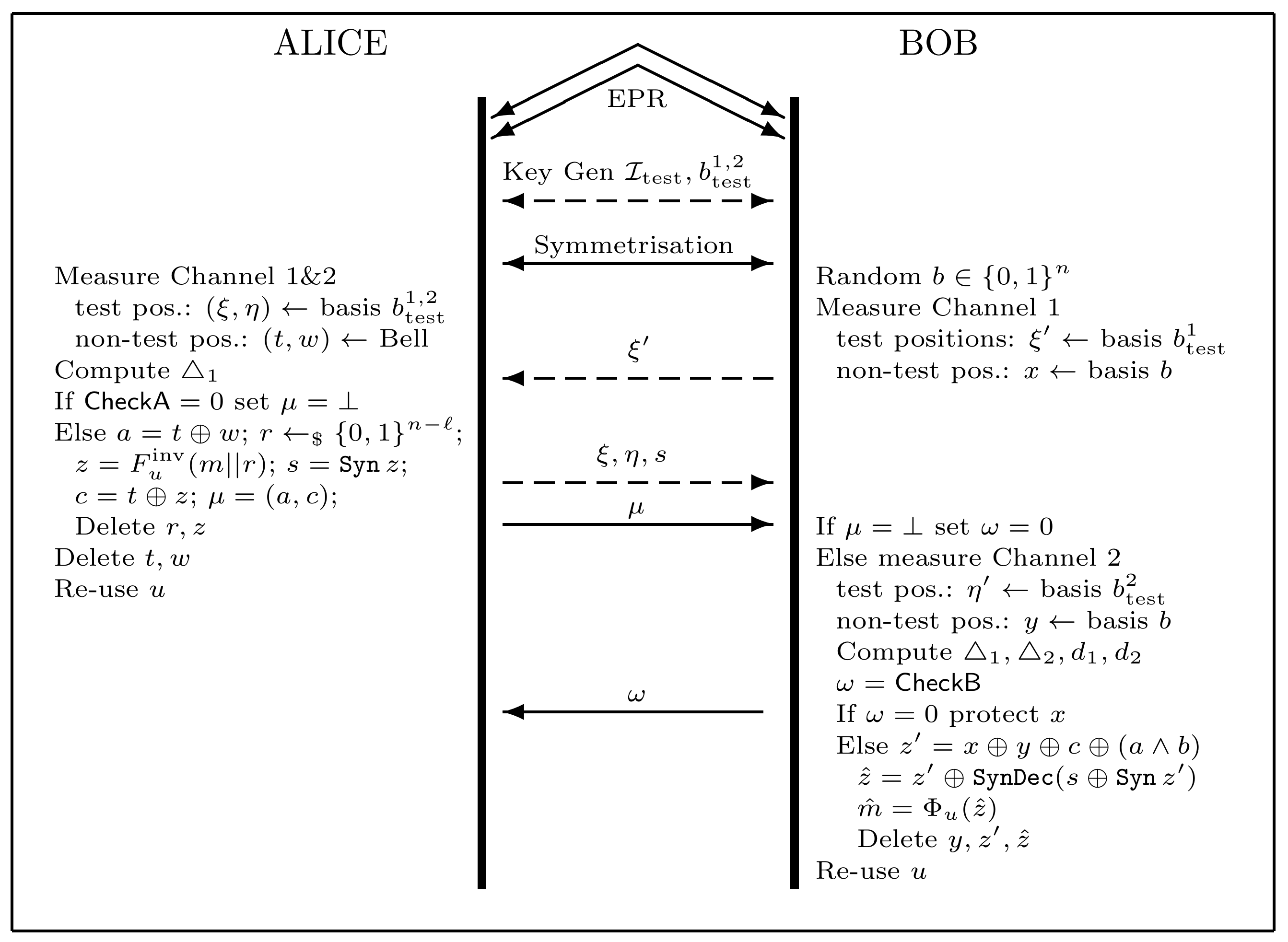}
\caption{\it 
Diagram of the EPR version of the protocol. 
A dashed line is a confidential classical channel.
}
\label{fig:EPR}
\end{figure}

The steps are shown in Figure \ref{fig:EPR} and listed below.
An untrusted source creates $2(n+\ql)$ noisy EPR pairs in the $\ket{\Phi_{00}}$ state, sends half of each pair to Alice and half to Bob. They perform the following actions:

\underline{\bf Key generation}:\\
Alice and Bob agree on new shared keys
$\cI_{\rm test}, b^{1,2}_{\rm test}$.
The key $u$ is re-used from the previous run.
Eve has no knowledge of these keys.

\underline{\bf Symmetrisation}:\\
Alice and Bob agree on a secret random permutation $\pi$
and two public random strings $\qa_1,\qa_2 \in \{0,1,2,3\}^{n+\ql}$.
They each apply $\pi$ to their own set of EPR qubits.
They each apply the Pauli operation 
$\qs(\qa_1)=\bigotimes_{i=1}^{n+\ql}\qs_{\qa_1[i]}$ 
on their own EPR qubits in Channel~1,
and $\qs(\qa_2)$ in Channel~2.
Then they forget $\qa_1,\qa_2$.

\underline{\bf Bob}:\\
Generate random $b\in\bits^n$.
Measure the qubits of Channel~1.
At the test positions $\cI_{\rm test}$, measure in basis $b_{\rm test}^1$. 
This yields outcome~$\xi'\in\bits^\ql$.
At the non-test positions measure in basis~$b$, yielding outcome $x\in\bits^n$.
Send $\xi'$ through the confidential channel.

\underline{\bf Alice}:\\
In the test positions $\cI_{\rm test}$, measure the Channel~1 qubits in basis $b_{\rm test}^1$,
and the Channel~2 qubits in basis $b_{\rm test}^2$. 
This yields strings $\xi$ and $\qh$ respectively.
In each non-test position measure the pair of qubits in the Bell basis $(\ket{\qF_{tw}})_{t,w\in\bits}$.
This yields outcome $t,w\in\bits^n$.
Compute $\tri_1$ from $\xi,\xi',b_{\rm test}^1$.
Perform ${\sf CheckA}(\tri_1)$.
If the result is~0, set $\mu=\bot$.
Otherwise \{Set $a=t\oplus w$;
Draw random $r\in\bits^{n-\ell}$; 
$z = F_u^{\rm inv}(m\|r)$;
$s = {\tt Syn}\, z$;
$c=t\oplus z$; 
$\mu=(a,c)$.
Send $\xi,\qh,s$ over the confidential channel;
Delete $r,z,s$.\}
Send $\mu$ over the authenticated channel.
Delete $t,w$.

\underline{\bf Bob}:\\
If $\mu=\bot$ set $\qo=0$.
If $\mu\neq\bot$ then\{
Measure the qubits of Channel~2.
In the test positions $\cI_{\rm test}$ measure in basis $b_{\rm test}^2$, yielding $\qh'\in\bits^\ql$.
In the non-test positions measure in basis $b$, yielding $y\in\bits^n$.
Compute $\tri_1,\tri_2,d_1,d_2$ from $\xi,\xi',\qh,\qh',b_{\rm test}^1,b_{\rm test}^2$.
Set $\qo={\sf CheckB}(\tri_1,\tri_2,d_1,d_2)$.
If $\qo=0$ protect~$x$. Else continue.
Compute $z' = x \oplus y \oplus c \oplus (b \land a)$;
$\hat z = z'\oplus{\tt SynDec}(s \oplus {\tt Syn}\, z')$;
$\hat m = \Phi_u(\hat z)$.
Delete $y,z',\hat z$.\}

Send $\qo$.

\section{Eve's state}
\label{sec:stateEve}
\label{sec:collectiveState}

We derive the state of Eve's ancilla.
This will be used at the end of Section~\ref{sec:secproof}.

\subsection{Effect of symmetrisation}

We consider the (noisy) state of the EPR particles received by Alice and Bob.
Without Post-selection we would have had to consider a general $2^{4(n+\ql)}$-dimensional state.
Post-selection instead allows us to restrict the analysis to collective attacks, which lead to a factorised state
which we write as 
$(\qs^{A_1 B_1 A_2 B_2})^{\otimes(n+\ql)}$.
Here `A' and `B' denote the subsystems of Alice and Bob, and the indices 1,2 denote the channel.
Note that $\qs^{A_1 B_1 A_2 B_2}$ is 16-dimensional.

Next we look at the effect of the noise symmetrisation as discussed in Section~\ref{sec:paulitrick}.
Averaging over the random Pauli operators yields a new state $\tilde \qs$ given by
\be
	\tilde \qs^{A_1B_1A_2B_2} = 
	\frac{1}{16} \sum_{\qa_1,\qa_2=0}^3 (\qs_{\qa_1} \otimes \qs_{\qa_1} \otimes \qs_{\qa_2} \otimes \qs_{\qa_2}) 
	\qs^{A_1B_1A_2B_2} (\qs_{\qa_1} \otimes \qs_{\qa_1} \otimes \qs_{\qa_2} \otimes \qs_{\qa_2}).
\ee
\begin{lemma}
\label{lem:diag}
$\tilde \qs^{A_1B_1A_2B_2}$ is diagonal in the basis 
$\Big(\ket{\Phi_{a_1 b_1}} \otimes \ket{\Phi_{a_2 b_2}}\Big)_{a_1,b_1,a_2,b_2\in\bits}$,
where $\ket{\qF_{ab}}$ stands for the Bell states as defined in Section~\ref{sec:notation}.
\end{lemma}
\underline{Proof}: 
For a general 16-dimensional mixed state we can write
\be
	\qs^{\rm A_1B_1A_2B_2} = \sum_{a_1 b_1 a_2 b_2} \sum_{a_1' b_1' a_2' b_2'} \nu^{a_1 b_1 a_2 b_2}_{a_1' b_1' a_2' b_2'} 
	\ket{\Phi_{a_1b_1}}\bra{\Phi_{a_1'b_1'}} \otimes \ket{\Phi_{a_2b_2}}\bra{\Phi_{a_2'b_2'}}.
\ee
In each channel separately the effect of the randomisation is
$\fr14 \sum_\qa (\qs_\qa \otimes \qs_\qa) \ket{\Phi_{ab}} \bra{\Phi_{a'b'}} (\qs_\qa \otimes \qs_\qa) =$
$\qd_{aa'}\qd_{bb'} \ket{\Phi_{ab}} \bra{\Phi_{a'b'}}$.
This yields 
$\tilde \qs^{A_1B_1A_2B_2} = \sum_{a_1 b_1 a_2 b_2}  \nu^{a_1 b_1 a_2 b_2}_{a_1 b_1 a_2 b_2}  
\ket{\Phi_{a_1b_1}}\bra{\Phi_{a_1b_1}} \otimes \ket{\Phi_{a_2b_2}}\bra{\Phi_{a_2 b_2}}$.
\hfill $\square$

We denote the diagonal elements of $\tilde \qs^{A_1B_1A_2B_2}$ as $\ql^{a_1b_1}_{a_2b_2}$ for $a_1,b_1,a_2,b_2 \in\bits$. 
Eve holds the purification of $\tilde \qs^{A_1B_1A_2B_2}$. 
The joint state of Alice Bob and Eve is described by a pure state $\ket{\psi^{\rm ABE}}$,
\bea
\ket{\psi^{\rm ABE}} &=& \sum_{a_1b_1a_2b_2} \sqrt{\ql^{a_1b_1}_{a_2b_2}} 
\ket{\Phi_{a_1b_1}}\otimes\ket{\Phi_{a_2b_2}}\otimes\ket{e^{a_1b_1}_{a_2b_2}}.
\eea
Here $\ket{e^{a_1b_1}_{a_2b_2}}$ is an orthonormal basis in Eve's 16-dimensional Hilbert space.

\subsection{Effect of channel monitoring}
\label{sec:EvestateMonitoring}

There are 15 degrees of freedom in $\tilde \qs^{A_1B_1A_2B_2}$.
However, the channel monitoring {\sf CheckA}, {\sf CheckB} imposes a large number of constraints.
For the bit error probability in Channel~$i$, as a function of the monitoring basis $b\in\{0,1,2\}$, we write
\be
	r_i(b)\isdef \sum_{x\in\bits}\bra{\qj^b_x}\otimes\bra{\qj^b_{\overline x\oplus\qd_{b2}}}\tilde\qs^{\rm A_i B_i}
	\ket{\qj^b_x}\otimes\ket{\qj^b_{\overline x \oplus\qd_{b2}}}
\label{constraintA}
\ee
where the $\qd_{b2}$ occurs because the $\qF_{00}$ Bell state has a bit flip in the $b=2$ basis ($y$ basis).
Similarly, by $s(b,b')$ we denote the probability that a flip is detected {\em both} Channels,
as a function of the monitoring basis $b$ in Channel~1 and $b'$ in Channel~2,
\be
	s(b,b')\isdef\sum_{p,q\in\bits}\bra{\qj^b_p}\otimes\bra{\qj^b_{\overline p\oplus\qd_{b2}}}
	\otimes \bra{\qj^{b'}_q}\otimes\bra{\qj^{b'}_{\overline q\oplus\qd_{b'2}}}
	\tilde\qs^{\rm A_1 B_1 A_2 B_2}
	\ket{\qj^b_p}\otimes\ket{\qj^b_{\overline p\oplus\qd_{b2}}}
	\otimes \ket{\qj^{b'}_q}\otimes\ket{\qj^{b'}_{\overline q\oplus\qd_{b'2}}}.
\ee
If only {\sf CheckA} is passed, the monitoring imposes that
$\exists_{\qb\leq\qb^*} \forall_{b\in\{0,1,2\}}\;\; r_1(b)\approx \qb$.
If {\sf CheckB} is passed, the channel monitoring imposes 15 constraints,
\be
	\exists_{\qb\leq\qb^*, \qg\leq\qg^*} \forall_{b,b'\in\{0,1,2\}} \quad 
	r_1(b)\approx\qb \;\wedge\; r_2(b')\approx\qg \;\wedge\; s(b,b')\approx\qb\qg.
\label{constraints15}
\ee
Since we will study asymptotics only, we will treat the `$\approx$' in (\ref{constraints15})
as equal signs.

\begin{lemma}
\label{lem:solNoAbort}
If ${\sf CheckA}=1$ then it holds that 
$\sum_{a_2b_2} \ql_{a_2b_2}^{a_1b_1} = \frac{\qb}{2} + \qd_{a_1,0}\qd_{b_1,0} (1-2\qb)$.
\end{lemma}
\underline{Proof}:
We have $\tilde\qs^{\rm A_1B_1}=\tr_{\rm \!A_2B_2}\tilde\qs^{\rm A_1B_1A_2B_2}$
$=\sum_{a_1b_1}(\sum_{a_2b_2} \ql^{a_1b_1}_{a_2b_2}) \ket{\Phi_{a_1b_1}}\bra{\Phi_{a_1b_1}}$.
We introduce abbreviated notation $c^{a_1b_1}=\sum_{a_2b_2} \ql^{a_1b_1}_{a_2b_2}$.
The $b=0$ constraint (\ref{constraints15}) in Channel~1 gives $\sum_{b_1}c^{1,b_1}=\qb$.
Siilarly, the $b=1$ constraint gives $\sum_{a_1}c^{a_1,1}=\qb$,
and the $b=2$ constraint gives $\sum_{a_1}c^{a_1\overline{a_1}}=\qb$.
Furthermore, normalisation of $\tilde\qs$ requires $\sum_{a_1b_1}c^{a_1b_1}=1$.
Together this uniquely fixes $c^{a_1b_1}=\frac{\qb}{2} + \overline{a_1}\overline{b_1} (1-2\qb)$.
\hfill $\square$

\begin{lemma}
\label{lem:solAccept}
If ${\sf CheckB}=1$ then $\ql^{a_1b_1}_{a_2b_2} = \big[\frac{\qb}{2}+\qd^{a_1b_1}_{0,0}(1-2\qb)\big]\big[\frac{\qg}{2}+\qd^{a_2b_2}_{0,0}(1-2\qg)\big]$.
\end{lemma}
\underline{Proof:}
{\sf CheckB} comprises {\sf CheckA}. Hence we can use the result of Lemma~\ref{lem:solNoAbort}.
Also, we get the equivalent of Lemma~\ref{lem:solNoAbort} for Channel~2,
$\sum_{a_1b_1} \ql_{a_2b_2}^{a_1b_1} = \frac{\qg}{2} + \qd_{0,0}^{a_2b_2} (1-2\qg)$.
The $s(b,b')$ constraints in (\ref{constraints15}) yield
$\sum_{a_1a_2} \ql^{a_11}_{a_21} = \sum_{a_1a_2} \ql^{a_11}_{a_2\bar a_2} = \sum_{a_1b_2} \ql^{a_11}_{1b_2} = \sum_{a_1a_2} \ql^{a_1\bar a_1}_{a_21} = \sum_{a_1a_2} \ql^{a_1\bar a_1}_{a_2\bar a_2}= \sum_{a_1b_2} \ql^{a_1\bar a_1}_{1b_2}= \sum_{b_1a_2} \ql^{1 b_1}_{a_21} = \sum_{b_1a_2} \ql^{1 b_1}_{a_2\bar a_2} = \sum_{b_1b_2} \ql^{1 b_1}_{1b_2} =\qb\qg$.
Solving this system of equations yields the claim.
\hfill $\square$

Lemma~\ref{lem:solAccept} tells us that
in the Accept case ($\qo={\sf CheckB}=1$),  Alice and Bob's state reduces to the tensor product of two states known from the 6-state analysis in \cite{RennerThesis}. 
The state depends only on the noise parameters $\qb$ and $\qg$.
\bea
\tilde \qs^{\rm A_1B_1A_2B_2}_{\tt accept} &=& \Big[ (1-\frac{3}{2}\qb) \ket{\Phi_{00}}\bra{\Phi_{00}}+\frac{\qb}{2} \big(\ket{\Phi_{01}}\bra{\Phi_{01}} + \ket{\Phi_{10}}\bra{\Phi_{10}} + \ket{\Phi_{11}}\bra{\Phi_{11}}\big)\Big] \otimes
\nn\\
&& \Big[ (1-\frac{3}{2}\qg) \ket{\Phi_{00}}\bra{\Phi_{00}}+\frac{\qg}{2} \big(\ket{\Phi_{01}}\bra{\Phi_{01}} + \ket{\Phi_{10}}\bra{\Phi_{10}} + \ket{\Phi_{11}}\bra{\Phi_{11}}\big)\Big].
\eea

\subsection{Conditioning on measurement outcomes}

Next we look at the effect of Alice and Bob's measurements in the non-test positions.
Alice performs a Bell measurement and Bob measures both his qubits in basis~$b$. 
Using the definition of the Bell states 
$\ket{\Phi_{t,a\oplus t}}_{A_1A_2} = \sum_{p\in\bits} \ket p_{A_1} \qs_x^t \qs_{z}^{a \oplus t} \ket p_{A_2}$ 
the measurements of Alice and Bob can be described by a single POVM.
At given $b\in\bits$ the joint measurement yields $x,y,a,t\in\bits$, 
\bea
	\cM^b_{xyat} &=& \ket{\phi^b_{xyat}}\bra{\phi^b_{xyat}}
	\\
	\ket{\phi^b_{xyat}}&=&\frac{1}{\sqrt2}\sum_p \ket{p} \ket{\qj^b_x} \qs_x^t\qs_z^{a\oplus t} \ket{p} \ket{\qj^b_y}.
\eea
Here $\forall_b \sum_{xyat}\cM^b_{xyat}={\mathbb 1}$.
Having a description of $\ket{\psi^{\rm ABE}}$ and $\ket{\phi_{bxyat}}$ allows us to write down Eve's state 
after the A and B subsystems have been measured, i.e. conditioned on the measurement outcomes.
At fixed $b\in\bits$ we write
\bea
	P_{xyat|b} & \isdef & \pr[xyat|b]
	\\
	P_{xyat|b} \cdot\qs^{\rm E}_{bxyat} &=&
	\tr_{\rm \!AB}\left(\ket{\qJ^{\rm ABE}} \bra{\qJ^{\rm ABE}} \cM^b_{xyat}\right)
	= \ket{\bar\qj^{\rm E}_{bxyat}} \bra{\bar\qj^{\rm E}_{bxyat}}
	\\
	\ket{\bar\qj^{\rm E}_{bxyat}} & \isdef & \inprod{\phi^b_{xyat}}{\qJ^{\rm ABE}}.
\eea
Here $\ket{\bar\qj^{\rm E}_{bxyat}}$ is a sub-normalised state with squared norm $P_{xyat|b}$.
We will write $\ket{\bar\qj^{\rm E}_{bxyat}}=\sqrt{P_{xyat|b}}\ket{\qj^{\rm E}_{bxyat}}$.

\begin{lemma}
\label{lem:psiEbar}
Eve's (sub-normalized) pure state conditioned on Alice and Bob's measurement outcomes is given by
\bea
\label{eq:psiEbar}
\ket{\bar \psi^{\rm E}_{bxyat}} &=& \frac1{2\sqrt{2}} \sum_{a_1b_1a_2b_2} \sqrt{\ql^{a_1b_1}_{a_2b_2}} (-1)^{b_2 t} \Big[ \bar b \qd_{x\oplus y \oplus t, a_1\oplus a_2} (-1)^{(b_1 + b_2 + a + t)(a_1 + x)}+\nn\\
&& b (-1)^{x a_1 + y(t+a_2)} \qd_{b_1\oplus b_2, x\oplus y\oplus a\oplus t}\Big] \ket{e^{a_1b_1}_{a_2b_2}}
\eea
\end{lemma}
\underline{Proof:} see Appendix \ref{sec:psiEbar}.\hfill $\square$\\

From Lemma~\ref{lem:psiEbar} we obtain
\be
	P_{xyat|b}=\inprod{\bar \psi^{\rm E}_{bxyat}}{\bar \psi^{\rm E}_{bxyat}}
	=\frac18\Big(\bar b \sum_{a_1b_1b_2} {\ql^{a_1b_1}_{a_1\oplus x\oplus y\oplus t,b_2}} 
	+ b \sum_{a_1b_1a_2} {\ql^{a_1b_1}_{a_2,b_1\oplus x\oplus y\oplus a\oplus t}}\Big).
\label{Pxyatgivenb}
\ee
\underline{Note}: 
filling in the $\ql$'s corresponding to an Accept (Lemma \ref{lem:solAccept}) gives 
\be
	P_{xyat|b}^{[\qo=1]}=\frac18 \big[\qd_{x\oplus y\oplus t\oplus ab,0} (1-\qb\star\qg) 
	+  \qd_{x\oplus y\oplus t\oplus ab,1} \qb\star\qg\big] 
\label{PxyatbAccept}
\ee
with the `$\star$' notation as defined in Section~\ref{sec:notation}.
This is as expected, since the bit error probability between $x$ and $y$
results from the concatenation of Channels 1 and~2.

\begin{corollary}
\label{corol:Pxatb}
It holds that $P_{xat|b}=\fr18$. 
\end{corollary}
\underline{\it Proof}:
$P_{xat|b}=\sum_y P_{xyat|b}$.
In (\ref{Pxyatgivenb}) the summation over $y$ gives rise to a full summation over
all 16 components of $\ql$. The Corollary follows from the normalisation
$\sum_{a_1 b_1 a_2 b_2}\ql^{a_1 b_1}_{a_2 b_2}=1$.
\hfill$\square$

The 16 states $\ket{\qj^{\rm E}_{bxyat}}$ (at fixed $b$) are {\em not} all mutually orthogonal,
though a subset is.
\begin{lemma}
\label{lem:orth}
\bea
\label{eq:orth}
\inprod{\psi^{\rm E}_{bxyat}}{\psi^{\rm E}_{b\overline x yat}} 
= \inprod{\psi^{\rm E}_{bxyat}}{\psi^{\rm E}_{b x\overline yat}} 
=\inprod{\psi^{\rm E}_{bxyat}}{\psi^{\rm E}_{bxya\overline t}} 
= \inprod{\psi^{\rm E}_{b x  ya t}}{\psi^{\rm E}_{b\overline x\,\overline ya\overline t}}  = 0.
\eea
\end{lemma}
\underline{Proof}: 
We take the inner product of two states of the form (\ref{eq:psiEbar}).
The cross terms contain $b\bar b$ and vanish.
The Kronecker deltas ensure that the inner product vanishes when an odd number out of the variables $\{x,y,t\}$ flip.
\hfill $\square$

{\bf Remark}.
Eve's overall state $\qs^{\rm E}_b=\EE_{xyat|b}\qs^{\rm E}_{bxyat}$ (at fixed $b$) is the purification
of $\tilde\qs^{\rm A_1B_1A_2B_2}$ and is therefore diagonal in the basis $\ket{e^{a_1b_1}_{a_2b_2}}$ and has
eigenvalues $\ql^{a_1b_1}_{a_2b_2}$. 
By way of consistency check we verify this by expanding
$\EE_{xyat|b}\qs^{\rm E}_{bxyat}=\sum_{xyat}P_{xyat|b}\ket{\qj^{\rm E}_{bxyat}}\bra{\qj^{\rm E}_{bxyat}}$
$=\sum_{xyat}\tr_{\!\rm AB}\ket{\qj^{\rm ABE}}\bra{\qj^{\rm ABE}}\cM^b_{xyat}$
$=\tr_{\!\rm AB}\ket{\qj^{\rm ABE}}\bra{\qj^{\rm ABE}}(\sum_{xyat}\cM^b_{xyat})$
$=\tr_{\!\rm AB}\ket{\qj^{\rm ABE}}\bra{\qj^{\rm ABE}}$,
which indeed produces the correct result.

Eve's state conditioned on the measurement outcomes {\em and Acceptance} is denoted as
$\qs^{\rm E}_{bxyat,\qo=1}$.

\begin{lemma}
\label{lem:eigAvgY}
Consider $\qs^{\rm E}_{bxat,\qo=1}=\EE_{y|bxat,\qo=1}\qs^{\rm E}_{bxyat,\qo=1}$.
The eigenvalues of $\qs_{bxat,\qo=1}^{\rm E}$ are $\qb \star \qg, 1-\qb\star\qg$ and fourteen times zero.
\end{lemma}
\underline{Proof}: 
We have $\qs_{bxat,\qo=1}^{\rm E} = 
\sum_y \frac{\pr[xyat|b,\qo=1]}{\pr[xat|b,\qo=1]} \qs_{bxyat,\qo=1}^{\rm E} 
= \sum_y 8 P^{[\qo=1]}_{xyat|b}\qs_{bxyat,\qo=1}^{\rm E}$.
Here we have used (\ref{PxyatbAccept}). 
From Lemma \ref{lem:orth} we know that 
$\qs_{bxyat,\qo=1}^{\rm E} \qs_{bx \overline y at,\qo=1}^{\rm E}=0$. 
Thus $\qs_{bxat,\qo=1}^{\rm E}$ is the weighted sum of two orthogonal projectors, with weights $\qb \star \qg$ and $1-\qb \star \qg$. 
\hfill $\square$

\begin{lemma}
\label{lem:eigAvgBXY}
Consider $\qs^{\rm E}_{bat}=\EE_{xy|bat}\qs^{\rm E}_{bxyat}$.
The spectral decomposition of $\qs_{bat}^{\rm E}$ is 
\bea
	\qs_{bat}^{\rm E} &= & \sum_{u,v\in\bits} \qL_{uv} \ket{V_{uv}^{at}} \bra{V_{uv}^{at}}
	\\
	\qL_{uv} &=& \sum_{k,\ell\in\bits} \ql^{k,\ell}_{k\oplus u,\ell\oplus v}
\label{defLambdauv}
	\\
	\ket{V_{uv}^{at}} &=& \sum_{k,\ell\in \bits} \sqrt{\lambda^{k\ell}_{k \oplus  u,\ell\oplus v}} \ket{e^{k\ell}_{k \oplus  u,\ell\oplus v}}
              (-1)^{(a+v)\bar k} (-1)^{t(k+\ell)}.
\eea
It holds that $\inprod{V^{at}_{uv}}{V^{at}_{u'v'}}=\qd_{uu'}\qd_{vv'}$.
\end{lemma}
\underline{Proof}: 
See Appendix~\ref{app:lembxy}.
\hfill $\square$

\section{Security proof}
\label{sec:secProof}

We prove the security of the modified protocol as presented in Section \ref{sec:modified}.
We show that the protocol is $\qe$-VSUE (Def.\,\ref{def:VSUE})
and $\qe$-KR (Def.\,\ref{def:KRinVSUE}), 
where $\qe$ decreases exponentially in $n$
if the message length $\ell$ is chosen appropriately.
In Section~\ref{sec:rate} we derive an expression for the asymptotic rate.

\subsection{CPTP maps}
\label{sec:CPTP}

We have the following identification between the abstract quantities 
in the security definitions on the one hand
and the protocol variables on the other hand. 

\begin{tabular}{|l|l|}
\hline
Definitions & EPR protocol
\\
Section~\ref{sec:securitydefs} & Section~\ref{sec:modified} \\
\hline
$m,\qo$ & $m,\qo$ \\
\hline
transcript $t$ & $b,s,a,c,\qa_1,\qa_2$
\\ \hline
$S_{\rm re}$ & u
\\ \hline
$S_{\rm once}$ & $\cI_{\rm test},b^{1,2}_{\rm test}$
\\ \hline
$P$ & $x$
\\ \hline
\end{tabular}

Here $b$ and $s$ are counted as part of the transcript because they eventually leak.\footnote{
$s$ leaks because its OTP eventually leaks.
}
The random Paulis $\qa_1,\qa_2$ will not appear explicitly in the analysis below because their symmetrising effect
has already been accounted for.
Similarly, the $\cI_{\rm test},b^{1,2}_{\rm test}$ do not feature in the analysis because
the channel monitoring has been accounted for in Section~\ref{sec:EvestateMonitoring}.

We denote the overall action of Alice and Bob running the whole protocol
as a CPTP map $\cE$ acting on the `AB' subsystem of the noisy EPR state $\qr^{\rm ABE}$. 
\be
	\cE(\qr^{\rm ABE}) 
	= \rho^{XMUBSAC \rm E}_{\tt accept} + \rho^X \otimes \rho^{M U B S A C \rm E}_{\tt reject}.
\label{Eoutput}
\ee
In the `output' state we have traced out all variables except those in the table above.
Note that we have isolated $\qr^X$ in the Reject case, since the attacker model states
that $x$ does not leak upon Reject.
The coupling between Eve's subsystem `E' and all the other variables occurs through the 
measurement variables $b,x,y,a,t\in\bits^n$. 
We write $\qr^{\rm E}_{bxyat}$ for Eve's state conditioned on $b,x,y,a,t$.

We derive an expression for (\ref{Eoutput}) by starting from the state that additionally contains the 
variables $r,\qo,t,y$ and the channel monitoring variables `$D$', which we then trace out.
We have 
$\qr^{XMUBSAC RD\qO TY\rm E}$
$=\EE_{mubr}\EE_{xyat|b}\EE_d\sum_{sc\qo}$ $\ket{xmubsacrd\qo ty}\bra{xmubsacrd\qo ty}$
$\otimes\qr^{\rm E}_{bxyat}$ 
$\qd_{s,{\tt Syn}(c\oplus t)}$
$\qd_{c,t\oplus F_u^{\rm inv}(m||r)}$
$\qd_{\qo,{\sf CheckB}(d)}$.
Here the $u,b,r$ are uniform. 
The distribution of $x,y,a,t,d$ is determined by the noise that Eve is causing.
We trace out $r,t,y,d,\qo$.
The expectation $\EE_d$ then acts only on the Kronecker delta that contains $d$, which yields
a factor $\EE_d \qd_{\qo,{\sf CheckB}(d)}=\pr[\qO=\qo]$;
similarly, from $\EE_r$ we get a factor 
$\EE_r \qd_{c,t\oplus F_u^{\rm inv}(m||r)}=2^{\ell-n}\qd_{m,\qF_u(c\oplus t)}$.
The expectation $\EE_{y|bxat}$ acting on $\qr^{\rm E}_{bxyat}$
gives $\qr^{\rm E}_{bxat}$ by definition.
The result is
\bea
	\rho^{XMUBSAC \rm E}_{\tt accept} 
	\!\!\!\! &=&  \!\!
	\pr[\qO=1]\EE_{mu}\EE_{bxa}\sum_{sc}
	\ket{xmubsac}\bra{xmubsac} \!\otimes \!\! \EE_{t|bxa} \!\! \qr^{\rm E}_{bxat}
	\qd_{s,{\tt Syn}(c\oplus t)} 2^{\ell-n}\qd_{m,\qF_u(c\oplus t)}
	\quad
\label{rhoaccept}
	\\
	\rho^{MUBSAC \rm E}_{\tt reject} 
	&=&
	\pr[\qO=0]\EE_{mu}\EE_{ba}\sum_{sc}
	\ket{mubsac}\bra{mubsac}\otimes \EE_{t|ba}\qr^{\rm E}_{bat}
	\qd_{s,{\tt Syn}(c\oplus t)} 2^{\ell-n}\qd_{m,\qF_u(c\oplus t)}
\label{rhoreject}
\eea
Next we determine the CPTP map $\cF$ that corresponds to the `ideal' functionality.
First,
in order to achieve $0$-KR according to Def.\,\ref{def:KRinVSUE}
the re-used key $u$ must decouple from all the other variables.
I.e. $\cF$ must be such that
$\cF(\qr^{\rm ABE})=\qr^U\otimes\tr_{\!U}\, \cE(\qr^{\rm ABE})$.
If we apply $\tr_{\!U}$ to (\ref{rhoaccept},\ref{rhoreject}) we see
the expression $\EE_u \qd_{m,\qF_u(c\oplus t)}$ appearing, which evaluates to $2^{-\ell}$
due to the fact that $\qF$ is a pairwise independent hash.
This causes $\EE_m\ket m\bra m$ to decouple from the rest of the state;
hence $0$-VSUE (Def.\,\ref{def:VSUE}) is also satisfied.
We get
\bea
	\cF(\qr^{\rm ABE}) &=& \qr^{MU}\otimes (\qr^{XBSAC\rm E}_{\tt accept} + \qr^X \otimes \rho^{BSAC\rm E}_{\tt reject})
\label{Foutput}
	\\
	\qr^{XBSAC\rm E}_{\tt accept} & = &
	\pr[\qO=1]\EE_{bxa}\sum_{sc}2^{-n}
	\ket{xbsac}\bra{xbsac}\otimes \EE_{t|bxa}\qr^{\rm E}_{bxat}
	\qd_{s,{\tt Syn}(c\oplus t)} 
	\\
	\rho^{BSAC\rm E}_{\tt reject} &=&
	\pr[\qO=0]\EE_{ba}\sum_{sc}2^{-n}
	\ket{bsac}\bra{bsac}\otimes \EE_{t|ba}\qr^{\rm E}_{bat}
	\qd_{s,{\tt Syn}(c\oplus t)} .
\eea
Finally we obtain an expression for $\|\cE-\cF\|_{\diamond}$
by taking the trace distance between (\ref{Eoutput}) and (\ref{Foutput}).
Using the triangle inequality to separate the Accept and Reject contribution, we get
\bea
	\|\cE-\cF\|_{\diamond} &\leq& D_{\rm acc}+D_{\rm rej}
\label{EFdiamondaccrej}
	\\
	D_{\rm acc} &=& \pr[\qO=1] \EE_{mu}\EE_{bxa}\sum_{sc}2^{-n}
	\Big\| \EE_{t|bxa}\qr^{\rm E}_{bxat}
	\qd_{s,{\tt Syn}(c\oplus t)} (2^\ell\qd_{m,\qF_u(c\oplus t)}-1)\Big\|_1
	\\
	D_{\rm rej} &=& 
	\pr[\qO=0]\EE_{mu}\EE_{ba}\sum_{sc}2^{-n}
	\Big\| \EE_{t|ba}\qr^{\rm E}_{bat}
	\qd_{s,{\tt Syn}(c\oplus t)} (2^\ell\qd_{m,\qF_u(c\oplus t)}-1)\Big\|_1 .
\eea
Note that a bound $\|\cE-\cF\|_{\diamond}\leq\qe$ implies $\qe$-VSUE  and $\qe$-KR.

\subsection{Main Theorem}
\begin{theorem}
\label{th:main}

Let $\cE$ be the CPTP map according to the protocol of Section~\ref{sec:modified},
and let $\cF$ be its idealized version that satisfies $0$-VSUE and $0$-KR
as defined in Defs.\;\ref{def:VSUE} and~\ref{def:KRinVSUE}.
It holds asymptotically that
\bea
	\|\cE-\cF\|_\diamond &\leq & \qe_{\rm acc}+\qe_{\rm rej}
	\\
\label{epsaccept}
	\qe_{\rm acc} &=& 
	\pr[\qO=1]\cdot\min\left(1,\sqrt{2^{\ell-n+nh(1-\frac{3}{2}\qb^*,\frac{\qb^*}{2},\frac{\qb^*}{2},\frac{\qb^*}{2}) 
	+ nh(1-\frac{3}{2}\qg^*,\frac{\qg^*}{2},\frac{\qg^*}{2},\frac{\qg^*}{2}) }}\;\right)
	\\
	\qe_{\rm rej} &=& \pr[\qO=0]\cdot
	\min\left(1,\sqrt{2^{\ell-n+nh(1-\frac32\qb^*,\frac{\qb^*}{2},\frac{\qb^*}{2},\frac{\qb^*}{2})+nh(\qb^*\star\qg^*) }} \right).
\label{epsreject}
\eea
\end{theorem}
Note that $\pr[\qO=1]$ is exponentially small in $n$ 
when $\qb>\qb*$ and/or  $\qg>\qg*$. (This follows from e.g.\;Hoeffding's inequality).

\subsection{Proof of Theorem \ref{th:main}}
\label{sec:secproof}

We start from (\ref{EFdiamondaccrej}).
Since $D_{\rm acc}$ and $D_{\rm rej}$ are both defined as a distance  between sub-normalised states,
we have $D_{\rm acc}\leq\pr[\qO=1]$ and $D_{\rm rej}\leq\pr[\qO=0]$.
We note that $D_{\rm acc}$ and $D_{\rm rej}$ are very similar expressions.
In order to treat them in one go we introduce the notation `$q$' which 
stands for $(b,x,a)$ in the Accept case and for $(b,a)$ in the Reject case.

Next we introduce smoothing as in \cite{RennerThesis}, i.e.\;we
consider states $\bar \rho$ that are $\qe$-close to $\rho$ in terms of trace distance.
Doing this incurs a penalty $\propto\qe$ in the diamond norm, but this penalty vanishes
asymptotically.
After these steps we can write the smoothened version of $D_{\rm acc}$,$D_{\rm rej}$ both in the form
$\pr[\qO=0/1]\bar D$.
\bea
	\bar D &\isdef& 
	\EE_{muscq}2^{n-\qk}
	\Big\| \EE_{t|q}\bar\qr^{\rm E}_{qt}
	\qd_{s,{\tt Syn}(c\oplus t)} (2^\ell\qd_{m,\qF_u(c\oplus t)}-1)\Big\|_1.
\eea
Here we have written, in slight abuse of notation, 
$\EE_s(\cdots)=2^{\qk-n}\sum_s(\cdots)$
and $\EE_c(\cdots)=2^{-n}\sum_c(\cdots)$.
We derive an upper bound on $\bar D$ using steps that are very similar to
the derivation of the Leftover Hash Lemma \cite{TSSR2011}.
We rewrite the 1-norm as the trace over a square root. 
Then we use Jensen's inequality to `pull' the $\EE_u$, $\EE_s$ and $\EE_q$ expectation into the square root.
Next we exploit the pairwise independence property of the hash function $\qF$, yielding 
a result that can be formulated in terms of smooth R\'{e}nyi entropies
$S_0^\qe$ and $S_2^\qe$.
Finally we substitute the factorised form of Eve's state (due to Postselection)
and make use of the limiting behaviour as described in Section~\ref{sec:smooth}
to obtain von Neumann entropies.

\bea
	\bar D \!\!\!\!\!\!\!\!\!\!\!\!&=& \!\!\!\!\!\!\!\!\!
	\EE_{muscq} \!\!\!\! 2^{n-\qk} \tr\! \sqrt{\EE_{t|q}\EE_{t'|q}\bar\qr^{\rm E}_{qt}\bar\qr^{\rm E}_{qt'}
	\qd_{s,{\tt Syn}(c \oplus t)} \qd_{s,{\tt Syn}(c \oplus t')} 
	(2^\ell\qd_{m,\qF_u(c\oplus t)}\!-\!1)(2^\ell\qd_{m,\qF_u(c\oplus t')}\!-\!1)}
	\quad
	\\
	&\stackrel{\rm {Jensen}}{\leq}& 
	\EE_{mc}2^{n-\qk} \tr \Big\{ \EE_{usq} \EE_{tt'|q} \bar \rho_{qt}^{\rm E}\bar \rho_{qt'}^{\rm E} 
	\qd_{s,{\tt Syn}(c\oplus t)}\qd_{s,{\tt Syn}(c\oplus t')}
	\\&&
	\big[ 2^{2\ell} \qd_{m,\Phi_u(c\oplus t)}\qd_{m, \Phi_u(c\oplus t')} - 2^\ell\qd_{m,\Phi_u(c\oplus t)} - 2^\ell\qd_{m,\Phi_u(c\oplus t')} + 1 \big] \Big\}^{\frac12}
\nn\\
&\stackrel{\rm pair.indep.}{=}& 
	\EE_{mc}2^{n-\qk} \tr \sqrt{ \EE_{qtt'} \bar \rho_{qt}\bar \rho_{qt'} [\EE_s \qd_{s,{\tt Syn}(c\oplus t)}\qd_{s,{\tt Syn}(c\oplus t')}] 2^{\ell} \qd_{tt'} }
\label{Eutrick}	
\\
&=& \sqrt{2^{n-\qk}} \tr \sqrt{\EE_{qtt'} \bar \rho^{\rm E}_{qt}\bar \rho^{\rm E}_{qt'} 2^\ell \qd_{tt'}}
\\&\stackrel{\rm {Jensen}}{\leq}& 
\sqrt{2^{n-\qk}} \tr \sqrt{\EE_{qt}\EE_{q't'} |\cQ|\qd_{qq'}2^\ell \qd_{tt'} \bar \rho^{\rm E}_{qt}\bar \rho^{\rm E}_{q't'} }\\
&=& \sqrt{|\cQ|\, 2^{n-\qk+\ell}}\tr_{\!\rm E} \sqrt{\tr_{\! QT} \Big(\bar \rho^{QT{\rm E}}\Big)^2}
\\&\stackrel{\rm {Jensen}}{\leq}& 
\sqrt{|\cQ|\, 2^{n-\qk+\ell}} \sqrt{{\rm rank}(\tr_{QT} (\bar \rho^{QT{\rm E}})^2)}\sqrt{\tr_{QT{\rm E}} (\bar \rho^{QT{\rm E}})^2)}
\\ &=& 
\sqrt{|\cQ|\,2^{n-\qk+\ell}\, {\rm rank}(\bar \rho^{\rm E}) \tr (\bar \rho^{QT{\rm E}})^2}
\label{ranktrick}
\\&=& 
\sqrt{|\cQ|\,2^{n-\qk+\ell} 2^{S_0(\bar \rho^E) - S_2(\bar \rho^{QT{\rm E}})}}
\\&=& 
\sqrt{|\cQ|\, 2^{n-\qk+\ell}2^{S^\qe_0(\rho^E) - S^\qe_2(\rho^{QT{\rm E}})}}
\\&\stackrel{\rm postselection}{=}& 
\sqrt{|\cQ|\, 2^{n-\qk+\ell}2^{S^\qe_0([\qs^{\rm E}]^{\otimes n}) - S^\qe_2([\qs^{QT\rm E}]^{\otimes n})}}
\\&\stackrel{\rm {asymptotic}}{\to}& 
\sqrt{|\cQ|\,2^{n-\qk+\ell}\, 2^{nS(\qs^{\rm E}) - nS(\qs^{QT\rm E})}}.
\label{eq:Dacc}
\eea
In (\ref{Eutrick}) we used $\EE_u \qd_{m,\qF_u(\cdots)}=2^{-\ell}$ and
$\EE_u 2^{2\ell} \qd_{m,\Phi_u(c\oplus t)}\qd_{m, \Phi_u(c\oplus t')} = 2^\ell \qd_{tt'} + (1-\qd_{tt'})$.
In (\ref{ranktrick}) use is made of 
the rank equality ${\rm rank} \big(\tr_{QT}(\bar \rho^{QTE})^2\big) = {\rm rank} (\tr \bar \rho^{\rm E})$, 
which was also used in \cite{RennerThesis}.
(A proof is written out in Appendix A of \cite{QKR_noise}.)
In (\ref{eq:Dacc}) the $\qs^{\rm E}$ and $\qs^{QT\rm E}$ are given by
\bea
	\qs^{\rm E}=\EE_{bxyat}\qs^{\rm E}_{bxyat} &;&
	\qs^{QT\rm E} = \EE_{qt}\ket{qt}\bra{qt}\otimes \qs^{\rm E}_{qt},
\eea
with $\qs^{\rm E}_{bxat}=\EE_{y|bxat}\qs^{\rm E}_{bxyat}$ and 
$\qs^{\rm E}_{bat}=\EE_{xy|bat}\qs^{\rm E}_{bxyat}$.
We have $S(\qs^{QT\rm E})=\sH(QT)+\EE_{qt}S(\qs^{\rm E}_{qt})$.
From Lemma~\ref{lem:eigAvgY} we know that the eigenvalues of $\qs^{\rm E}_{bxat}$
do not depend on $bxat$ in the Accept case (which is exactly the case at hand).
Similarly, from Lemma~\ref{lem:eigAvgBXY} we see that the eigenvalues of
$\qs^{\rm E}_{bat}$ do not depend on $bat$.
Hence we can write $\EE_{qt}S(\qs^{\rm E}_{qt})=S(\qs^{\rm E}_{qt})$
where in the last expression the $q,t$ has arbitrary value.
Furthermore, from Corollary~\ref{corol:Pxatb} we get $\sH(BXAT)=4$
and $\sH(BAT)=3$.
Asymptotically it holds that $n-\qk\to nh(\qb^*\star\qg^*)$, since the error correcting code
is designed to deal with bit error rate $\qb^*\star\qg^*$, which results from the serial concatenation 
of noisy channels with error probability $\qb^*$ and~$\qg^*$.
Substitution into (\ref{eq:Dacc}) gives the following asymptotic result
\bea
	D_{\rm acc} &\leq&  \pr[\qO=1]\sqrt{2^{\ell-n+nh(\qb^*\star\qg^*)} 2^{nS(\qs^{\rm E}_{\qo=1})-nS(\qs^{\rm E}_{bxat,\qo=1})}}
\label{Daccbound}
	\\
	D_{\rm rej} &\leq&  \pr[\qO=0]\sqrt{2^{\ell-n+nh(\qb^*\star\qg^*)} 2^{nS(\qs^{\rm E})-nS(\qs^{\rm E}_{bat})}}.
\label{Drejbound}
\eea
\underline{Accept case}.
From Lemma~\ref{lem:solAccept} it follows that
$S(\qs^{\rm E}_{\qo=1})=
h(1-\fr32\qb,\fr\qb 2,\fr\qb 2,\fr\qb 2) + h(1-\fr32\qg,\fr\qg 2,\fr\qg 2,\fr\qg 2)$.
From Lemma~\ref{lem:eigAvgY} we get 
$S(\qs^{\rm E}_{bxat,\qo=1})=h(\qb\star\qg)$.
In the Accept case we know that $\qb\leq\qb^*$, $\qg\leq\qg^*$.
Substitution of these von Neumann entropies into (\ref{Daccbound}) yields~(\ref{epsaccept}).

\underline{Reject case}.
From Lemma~\ref{lem:eigAvgBXY} we have
\be
	S(\qs^{\rm E}) - S(\qs^{\rm E}_{bat}) = 
	\sum_{a_1b_1a_2b_2} \ql^{a_1b_1}_{a_2b_2} \log \frac{1}{\ql^{a_1b_1}_{a_2b_2}} + \sum_{uv} \qL_{uv} \log \qL_{uv}.
\label{SEminusSEbat}
\ee
We need to upper bound this expression under the four constraints specified in
Lemma~\ref{lem:solNoAbort}.
We use Lagrange optimisation, with constraint multipliers $\mu_{a_1b_1}$.
The Lagrangian is
\be
	\cL = \sum_{\substack{a_1b_1\\a_2b_2}}\ql^{a_1b_1}_{a_2b_2} \ln \frac{1}{\ql^{a_1b_1}_{a_2b_2}} + \sum_{uv} \qL_{uv} \ln \qL_{uv} + \sum_{a_1b_1} \mu_{a_1b_1}\big( \sum_{a_2b_2} \ql^{a_1b_1}_{a_2b_2} - \frac{\qb}{2}-\qd^{a_1,b_1}_{0,0}(1-2\qb)\big).
\ee
Computing the derivatives with respect to the $\ql$-parameters, and setting these derivatives to zero yields,
after some algebra,
\be
	\ql^{a_1b_1}_{a_2b_2}=\qL_{a_1\oplus a_2,b_1\oplus b_2} \exp \mu_{a_1 b_1}.
\label{Lderviv}
\ee
Summing (\ref{Lderviv}) over $a_2,b_2$  and using the constraints of Lemma~\ref{lem:solNoAbort} we solve for 
$\mu_{a_1b_1}$ and find $\exp \mu_{a_1b_1} = \frac{\qb}{2} + \qd^{a_1,b_1}_{0,0}(1-2\qb)$. 
Substituting this back into (\ref{Lderviv}) and using the definition of $\qL$ (\ref{defLambdauv}) 
we get a system of linear equations,
\be
	\ql^{a_1b_1}_{a_2b_2}=\left[\frac{\qb}{2} + \qd^{a_1,b_1}_{0,0}(1-2\qb)\right]\sum_{k\ell}
	\ql^{k\ell}_{k\oplus a_1\oplus a_2, \ell\oplus b_1\oplus b_2}.
\ee
The solution is
$\ql_{u\bar v}^{01}=\ql_{\bar u v}^{10} = \ql_{\bar u \bar v}^{11} = \ql^{00}_{uv} \frac{\qb/2}{1-3\qb/2}$
$\wedge \sum_{uv}\ql^{00}_{uv}=1-\fr32\qb$.
Then the simple relation $\qL_{uv}=(1-\fr32\qb)^{-1}\ql^{00}_{uv}$ holds.
This solution does not entirely fix all the parameters (three degrees of freedom are still open), but 
it does entirely fix (\ref{SEminusSEbat}),
\be
	S(\qs^{\rm E}) - S(\qs^{\rm E}_{bat}) \leq h(1-\frac32\qb,\frac{\qb}{2},\frac{\qb}{2},\frac{\qb}{2}).
\label{SdiffReject}
\ee
Because of the succeeded {\sf CheckA} we have $\qb\leq\qb^*$.
Finally, substitution of (\ref{SdiffReject}) with $\qb\leq\qb^*$ 
into (\ref{Drejbound}) yields (\ref{epsreject}).
\hfill$\square$

\underline{Remark:} An alternative way of deriving the {\em Reject} case result 
(\ref{SdiffReject})
would have been to consider a modification to the original protocol where Eve receives the basis choice $b$ just before Alice sends qubits. 
Eve can then measure $y=x\oplus t$ with perfect accuracy.
What remains for Eve is get information about $x$ from her $4$-dimensional ancilla state.
This is exactly the six-state QKD analysis. 

\subsection{Achievable asymptotic rate}
\label{sec:rate}

\begin{figure}[h]
\begin{center}
\includegraphics[width=.6\textwidth]{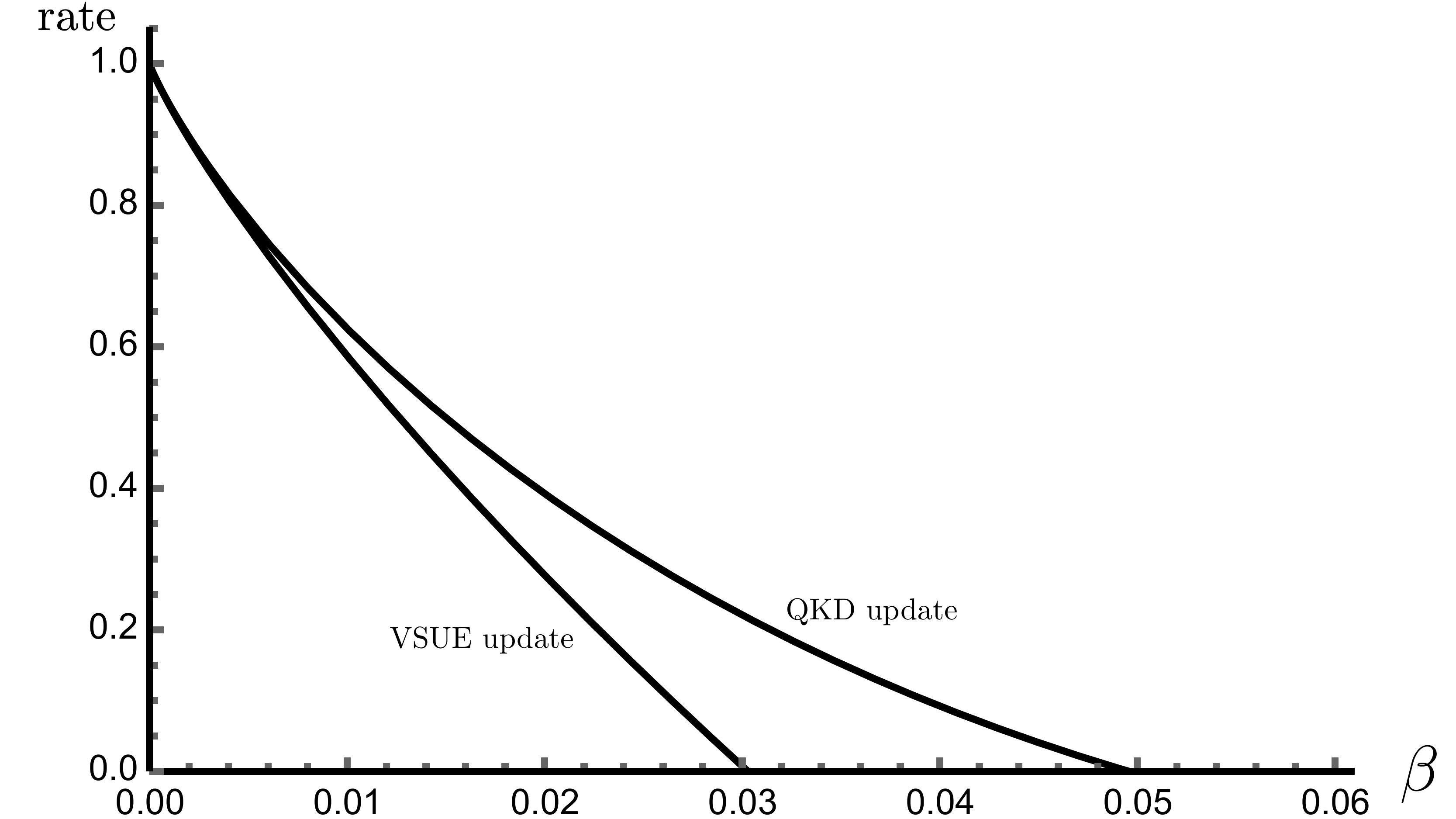}
\caption{\it Communication rate of our scheme with two different ways to update the single-use keys: through VSUE itself
and via 6-state QKD. 
The bit error rate on all channels is taken to be $\qb$.}
\label{fig:rates}
\end{center}
\end{figure}

Theorem~\ref{th:main} tells us how the length $\ell$ must be set 
so as to ensure an exponentially small diamond distance $\| \cE-\cF \|_\diamond$,
\be
	\ell\leq n- n h(1-\frac{3}{2}\qb^*,\frac{\qb^*}{2},\frac{\qb^*}{2},\frac{\qb^*}{2})
	-n \max\Big( h(\qb^*\star\qg^*), \;
	h(1-\frac{3}{2}\qg^*,\frac{\qg^*}{2},\frac{\qg^*}{2},\frac{\qg^*}{2}) \Big).
\ee
When $\qg^*$ is close to $\qb^*$, the second term in the max$()$ is dominant.
From this point on we set $\qg^*=\qb^*$. 
The requirement on $\ell$ becomes
\be
	\mbox{In case }\qg^*=\qb^*:\quad
	\ell \leq n-2n h(1-\frac{3}{2}\qb^*,\frac{\qb^*}{2},\frac{\qb^*}{2},\frac{\qb^*}{2}).
\ee
We define the {\bf rate} of a scheme as the length of the actual message sent, divided by the expended number of qubits.
Our scheme transmits a string of length $\ell$ while expending $n+\ql$ qubits, with $\ql\ll n$.
However, that is not the whole story, since the single-use keys 
$k_{\rm syn},k_{\rm test},\cI_{\rm test},b^{1,2}_{\rm test},\xi,\qh$ need to be refreshed somehow.
Of these keys, $k_{\rm syn}$ has length $nh(\qb^*\star\qb^*)$
while the others are at most of order $\log n$. 
One way of refreshing the keys is to send them as part of the message~$m$.
This reduces the actual message size from $\ell$ to $\ell -nh(\qb^*\star\qb^*)$, which results in the following rate,
\be
	\mbox{Refresh through VSUE}:\quad
	{\rm rate}= 1- 2h(1-\frac{3}{2}\qb^*,\frac{\qb^*}{2},\frac{\qb^*}{2},\frac{\qb^*}{2})-h(\qb^*\star\qb^*).
\label{rateViaVSUE}
\ee
Using VSUE to transport the single-use keys is overkill, since they are assumed to leak afterward.
It is much more efficient to do the refresh via QKD.
The rate of six-state QKD is $R_{\rm QKD}=1-h(1-\frac{3}{2}\qb^*,\frac{\qb^*}{2},\frac{\qb^*}{2},\frac{\qb^*}{2})$;
the number of qubits spent on transporting $nh(\qb^*\star\qb^*)$ bits is
$N_{\rm QKD}=nh(\qb^*\star\qb^*)/R_{\rm QKD}$;
the rate is $\ell/(n+N_{\rm QKD})$,
\be
	\mbox{Refresh by 6-state QKD}:\quad
	{\rm rate} = \frac{(1-2J)(1-J)}{1-J+h(\qb^*\star\qb^*)},
	\quad J\isdef h(1-\frac{3}{2}\qb^*,\frac{\qb^*}{2},\frac{\qb^*}{2},\frac{\qb^*}{2}).
\label{rateViaQKD}
\ee
The rates (\ref{rateViaVSUE},\ref{rateViaQKD}) are plotted in Fig.\,\ref{fig:rates}.
Key update via QKD is clearly the best option.
Note that these rates are significantly lower than
what can be achieved with a UE scheme that uses the quantum channel in a single direction
\cite{KRUE}. (In UE a positive rate is possible up to $\qb\approx0.12$.)

\clearpage
\section{Two-way Quantum Key Distribution}
\label{sec:QKD}

Our VSUE scheme can be modified such that it becomes a QKD scheme
with two-way use of the quantum channel, like \cite{BLMR2013}.
The modifications with respect to Section~\ref{sec:steps} are as follows.\\
(i) The message $m$ is replaced by a uniform random string, which serves as the QKD key.\\
(ii) Alice chooses $u$. (It is no longer a shared key). Bob confirms that he has received qubits. Then Alice reveals $u$.\\ 
(iii) The syndrome $s$ is sent in the clear, together with~$u$. 
$k_{\rm syn}$ does not exist.\\
(iv) The keys $b, k_{\rm test}, \cI_{\rm test}, b_{\rm test}^1,b_{\rm test}^2, \xi, \qh$ are kept secret forever.

The quantity to be upperbounded is
$D_{\rm QKD}=\| \qr_{\tt accept}^{MUSAC \rm E} - \qr^M \otimes \rho_{\tt accept}^{USAC \rm E} \|_1$,
since only the message needs to be kept safe, and it is only endangered in case of Accept.
Following the same steps as in Section \ref{sec:secproof} we find
\bea
\label{eq:secDefQKD}
	D_{\rm QKD}&=&
	\pr[\qO=1] 2^{-n}\EE_{mua}\sum_{sc} \Big\| \EE_{t|a}\rho_{at}^{\rm E} 
	\qd_{s,{\tt syn}(c \oplus t)} (2^\ell \qd_{m,\Phi_u(c \oplus t)} - 1) \Big\|_1
	\\ &\leq& 
	\pr[\qO=1]\cdot\min\Big(1,
	\sqrt{2^{\ell-n+ nh(\qb \star \qg)}2^{nS(\qs^{\rm E}_{\qo=1}) - n S(\qs_{at,\qo=1}^{\rm E})}} \Big).
\eea
From Lemma~\ref{lem:solAccept} we get
$S(\qs^{\rm E}_{\qo=1})=J_\qb+J_\qg$, where we use shorthand notation
$J_\qb\isdef h(1-\fr32\qb,\fr\qb2,\fr\qb2,\fr\qb2)$.
Furthermore, substitution of Lemma~\ref{lem:solAccept} into Lemma~\ref{lem:eigAvgBXY}
shows that the eigenvalues of $\qs^{\rm E}_{at,\qo=1}$ are
$\qL_{01}=\qL_{10}=\qL_{11}=\fr12\qb\star\qg$ and $\qL_{00}=1-\fr32\qb\star\qg$,
which yields 
$S(\qs^{\rm E}_{at,\qo=1})=J_{\qb\star\qg}$.
Hence we obtain
\be
	D_{\rm QKD} \leq
	\pr[\qO=1]\cdot\min\Big(1,
	\sqrt{2^{\ell-n+ nh(\qb \star \qg)}2^{nJ_\qb+nJ_\qg - n J_{\qb\star\qg}}} \Big).
\label{DQKDbound}
\ee
From (\ref{DQKDbound}) we see that $\ell$ has to be chosen as
$\ell\leq n-nh(\qb\star\qg)-nJ_\qb-nJ_\qg+nJ_{\qb\star\qg}$.
The corresponding rate is
\be
	\mbox{Key rate} = 1-h(\qb\star\qg)-J_\qb-J_\qg+J_{\qb\star\qg}.
\label{ourQKDrate}
\ee
Note that now there is no additional penalty from a syndrome mask,
since the syndrome is sent in the clear.

\begin{figure}
\begin{center}
\includegraphics[width=.6\textwidth]{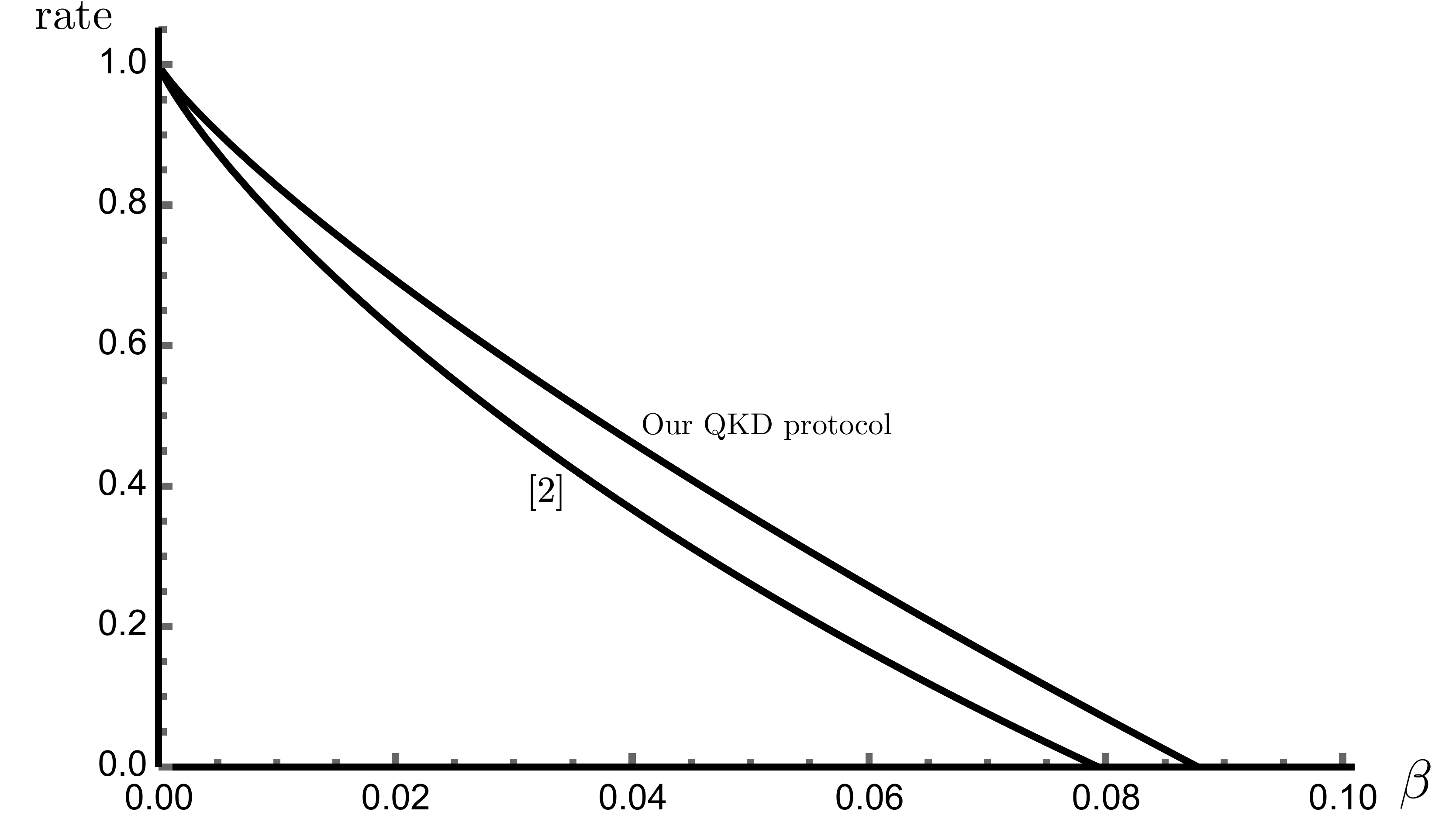}
\caption{\it 
Asymptotic key rate of two-way QKD as a function of the bit error rate $\qb$ on the quantum channels.
(The error rate of the two channels is set to be equal.)
Upper curve: our result (\ref{ourQKDrate}).
Lower curve:  \cite{BLMR2013}.
}
\label{fig:rateQKD}
\end{center}
\end{figure}

The authors of \cite{BLMR2013} used an entropic uncertainty relation to
prove achievable rate $1 - h(\qb \star \qg)- \min \big(h(\qb),h(\qg)\big)$ for LM05,  in the case of independent channel noise. 
To our knowledge that is the best known rate so far for a two-way version of QKD.
Our proof yields a higher rate (\ref{ourQKDrate}) when $\qg$ is close to~$\qb$. 
Fig.\,\ref{fig:rateQKD} shows a comparison of the rates for $\qb = \qg$.
This rate improvement could be due to the 6-state channel monitoring 
in combination with a proof technique similar to \cite{RennerThesis} which is able to exploit
that kind of monitoring.

\section{Discussion}
\label{sec:discussion}

We have constructed an Unclonable Encryption scheme with the 
additional property that even after a {\em Reject}
Alice is allowed to leak all her keys.
The price we pay for having VSUE, as compared to merely UE, is a lower rate
when there is channel noise;
this is due to the accumulation of noise from two channel uses instead of one.

The size of the shared keys is $2n+nh(\qb\star\qg)+\cO(\log n)$, 
where $u\in\bits^{2n}$ is re-usable and all the other keys have to be refreshed.
The optimal way to refresh is to use 6-state QKD. 
The key $k_{\rm syn}\in\bits^{nh(\qb\star\qg)}$ causes the main burden here.
It would be interesting to see if alternative ways of handling the leakage from error correction,
such as \cite{DS2005}, 
can help to improve the rate. 
This is left for future work.

We have used a proof technique with Postselection and random Pauli operators
because this technique allows us to prove a higher rate than techniques based on
entropic inequalities.
A slight disadvantage of this technique is that 
three-basis channel monitoring must become part of the protocol.

Our attacker model assumes that the noise on Channels 1 and~2 is independent.
In some circumstances it may be argued that the noise is {\em dependent} \cite{BLMR2013},
making it possible to achieve higher rates.
This is left as a topic for future work.

Finally we mention that schemes in which a message is encoded directly into
qubits are very sensitive to particle loss (erasures).
In our protocol, erasures in Channel~1 are harmless since Alice can ignore the erased positions.
Erasures in Channel~2 however are problematic; they force Alice and Bob to adapt their error-correcting
code to cope with erasures, which is costly.
A naive attempt to fix the problem would be to let Bob report erasure locations before Alice computes~$\mu$.
However, that would force Alice as well as Bob to wait for a response from the other party.
According to the attacker model, some of their variables would then become long-term secrets
and the VSUE property would be lost. 

\subsection*{Acknowledgements}
Part of this research was funded by NWO (CHIST-ERA project 651.002.003, ID\underline{\;\,}IOT).

\appendix

\section{Proof of Lemma~\ref{lem:psiEbar}: Eve's sub-normalized pure state}
\label{sec:psiEbar}
We write out $\ket{\bar \psi^{\rm E}_{bxyat}}$ as follows,
\bea
\label{eq:psiEbar2}
	\ket{\bar \psi^{\rm E}_{bxyat}} 
	&=& \inprod{\phi_{bxyat}}{\psi^{\rm ABE}}
	\\&=& 
	\frac1{\sqrt{2}} \sum_{pa_1b_1a_2b_2} \hskip-4mm
	\sqrt{\ql^{a_1b_1}_{a_2b_2}}\bra{p} {\bra{\qj^b_x}} \bra{p}{\bra{\qj^b_y}} (\one_2 \otimes \qs_x^{a_1}\qs_z^{b_1}
	\otimes \qs_z^{t\oplus a}\qs_x^t \otimes \qs_x^{a_2}\qs_z^{b_2}\otimes \one_{16})
	\\&& 
	\ket{\Phi_{00}}\ket{\Phi_{00}}\ket{e^{a_1b_1}_{a_2b_2}}
	\nn\\ &=&
	\frac1{2\sqrt{2}} \sum_{pqa_1b_1a_2b_2} \sqrt{\ql^{a_1b_1}_{a_2b_2}}  {\bra{\qj^b_x}}  \qs_x^{a_1}\qs_z^{b_1} \ket{p} \bra{p}\qs_z^{t\oplus a}\qs_x^t\ket{q}  {\bra{\qj^b_y}}  \qs_x^{a_2}\qs_z^{b_2} \ket{q} \ket{e^{a_1b_1}_{a_2b_2}}\\
&=&\frac1{2\sqrt{2}} \sum_{\substack{a_1b_1a_2b_2\\pq}}\hskip-3mm\sqrt{\ql^{a_1b_1}_{a_2b_2}} (-1)^{b_1 p + b_2 q} \big[ \qd_{b 0} \qd_{x,p\oplus a_1} \qd_{y,q\oplus a_2}+\frac{\qd_{b 1}}{2} (-1)^{x(p\oplus a_1)+y(q\oplus a_2)}\big]
	\\&&
	\big[\qd_{p 0} \qd_{t q} + (-1)^{a \oplus t} \qd_{t \bar q} \qd_{p 1} \big] \ket{e^{a_1b_1}_{a_2b_2}}
	\nn
\eea
\bea
	&=&
	\frac1{2\sqrt{2}}  \sum_{a_1b_1a_2b_2} \hskip-3mm \sqrt{\ql^{a_1b_1}_{a_2b_2}} (-1)^{b_2 t} \Big[ \qd_{b0} \big(\qd_{x a_1} \qd_{y \oplus t,a_2} + (-1)^{b_1 + b_2 + a + t} \qd_{x \bar a_1} \qd_{y \oplus t, \bar a_2}\big)+\\
&& \frac{\qd_{b1}}{2} (-1)^{x a_1 + y(t+a_2)}\big( 1+(-1)^{b_1+b_2 + x+y+a+t}\big)\Big] \ket{e^{a_1b_1}_{a_2b_2}}
\nn\\
	&=& 
	\frac1{2\sqrt{2}} \sum_{a_1b_1a_2b_2} \sqrt{\ql^{a_1b_1}_{a_2b_2}} (-1)^{b_2 t} \Big[ \bar b \qd_{x\oplus y \oplus t, a_1\oplus a_2} (-1)^{(b_1 + b_2 + a + t)(a_1 + x)}+\\
&& b (-1)^{x a_1 + y(t+a_2)} \qd_{b_1\oplus b_2, x\oplus y\oplus a\oplus t}\Big] \ket{e^{a_1b_1}_{a_2b_2}}\nn
\eea
where the last line is the claim. 
\hfill $\square$

\section{Proof of Lemma~\ref{lem:eigAvgBXY}}
\label{app:lembxy}

We have 
$\qs^{\rm E}_{bat}=\sum_{xy} \pr[xy|bat] \qs^{\rm E}_{bxyat}$ 
$= \sum_{xy}\frac{\pr[xyat|b]}{\pr[at|b]} \qs^{\rm E}_{bxyat}$ 
$= 4 \sum_{xy}  \ket{\bar \psi^{\rm E}_{bxyat}}\bra{\bar \psi_{bxyat}^{\rm E}}$. 
Next we check that the given $\ket{V^{at}_{uv}}$ are indeed eigenvectors.
Keeping track of the phases we get:
\bea
	\inprod{\bar \psi_{bxyat}^E}{V_{uv}^{at}} &=& 
	\frac{1}{2\sqrt2} \sum_{k \ell} \ql_{k\oplus u,\ell \oplus v}^{k,\ell} (-1)^{(a+v)\bar k+(k+v)t}
	\nn\\&& 
	\big[\bar b \qd_{x\oplus y \oplus t, u} (-1)^{(v+a+t)(k+x)} + b \qd_{x\oplus y\oplus a \oplus t,v} (-1)^{xk+(t+k+u)y} \big]
\eea
\bea
\qs_{bat}^E \ket{V_{uv}^{at}} &=& \frac{1}{2} \sum_{x y k \ell} \sum_{a_1b_1a_2b_2} \ql_{k\oplus u,\ell \oplus v}^{k,\ell} \sqrt{\ql^{a_1b_1}_{a_2b_2}}\ket{e^{a_1b_1}_{a_2b_2}}(-1)^{(a+v)\bar k+(k+v+b_2)t}
\nn\\&& \big[\bar b \qd_{x\oplus y \oplus t, u}\qd_{u,a_1\oplus a_2} (-1)^{(v+a+t)(k+x)+(b_1+b_2+a+t)(a_1+x)} 
\nn\\&&+ b \qd_{x\oplus y\oplus a \oplus t,v}\qd_{v,b_1\oplus b_2} (-1)^{xk+(t+k+u)y+xa_1+(t+a_2)y} \big]\\
&=& \frac{1}{2} \sum_{k \ell} \sum_{a_1b_1a_2b_2} \ql_{k\oplus u,\ell \oplus v}^{k,\ell} \sqrt{\ql^{a_1b_1}_{a_2b_2}}\ket{e^{a_1b_1}_{a_2b_2}}(-1)^{(a+v)\bar k+(k+v+b_2)t}
\nn\\&& \big[\bar b \qd_{u,a_1\oplus a_2} (-1)^{(v+a+t)k+(b_1+b_2+a+t)a_1} \sum_x (-1)^{(v+b_1+b_2)x}
\nn\\&&+ b \qd_{v,b_1\oplus b_2} (-1)^{(a+t+v)k+(a+t+v)a_1} \sum_y (-1)^{(a_1+a_2+u)y} \big]\\
&=& \sum_{k \ell} \sum_{a_1b_1} \ql_{k\oplus u,\ell \oplus v}^{k,\ell} \sqrt{\ql^{a_1b_1}_{a_1\oplus u,b_1 \oplus v}}\ket{e^{a_1b_1}_{a_1\oplus u,b_1 \oplus v}}(-1)^{(a+v)\bar k+(k+b_1)t}
\nn\\&& \big[\bar b (-1)^{(v+a+t)k+(v+a+t)a_1}+ b (-1)^{(a+t+v)k+(a+t+v)a_1} \big]
\\&=& \sum_{k\ell} \ql_{k\oplus u,\ell \oplus v}^{k,\ell} \sum_{a_1b_1} \sqrt{\ql^{a_1b_1}_{a_1\oplus u,b_1 \oplus v}}\ket{e^{a_1b_1}_{a_1\oplus u,b_1 \oplus v}} (-1)^{(a+v)\bar a_1+(a_1+b_1)t}.
\eea
We recognize $\qL_{uv} \ket{V_{uv}^{at}}$ with renamed summation variables.


\bibliography{archive_VSUE}

\end{document}